\documentclass[aps, prl, superscriptaddress, preprintnumbers, floatfix, longbibliography, twocolumn, 10pt]{revtex4-2}

\usepackage{graphicx,amsfonts,amsmath,amssymb,amstext}
\usepackage{float,wrapfig}
\usepackage{subfigure, psfrag}
\usepackage{dsfont}
\usepackage{xcolor}
\usepackage[colorlinks=true,urlcolor=blue,citecolor=blue,linkcolor=blue]{hyperref}
\usepackage{bm}

\newcommand{\be}{\begin{equation}}
\newcommand{\ee}{\end{equation}}
\newcommand{\ba}{\begin{eqnarray}}
\newcommand{\ea}{\end{eqnarray}}

\definecolor{purple}{rgb}{0.8,0,0.6}
\definecolor{darkgreen}{rgb}{0.00,0.6,0.00}

\extrafloats{100}

\begin{document}

\title{Entropy Wave Instability in Dirac and Weyl Semimetals}
\date{October 21, 2021}

\author{P.~O.~Sukhachov}
\email{pavlo.sukhachov@yale.edu}
\affiliation{Department of Physics, Yale University, New Haven, Connecticut 06520, USA}

\author{E. V. Gorbar}
\affiliation{Department of Physics, Taras Shevchenko National University of Kyiv, Kyiv, 03022, Ukraine}
\affiliation{Bogolyubov Institute for Theoretical Physics, Kyiv, 03143, Ukraine}

\author{I.~A.~Shovkovy}
\affiliation{College of Integrative Sciences and Arts, Arizona State University, Mesa, Arizona 85212, USA}
\affiliation{Department of Physics, Arizona State University, Tempe, Arizona 85287, USA}

\begin{abstract}
Hydrodynamic instabilities driven by a direct current are analyzed in 2D and 3D relativisticlike systems with the Dyakonov-Shur boundary conditions supplemented by a boundary condition for temperature. Besides the conventional Dyakonov-Shur instability for plasmons, we find an entropy wave instability in both 2D and 3D systems. The entropy wave instability is a manifestation of the relativisticlike nature of electron quasiparticles and a nontrivial role of the energy current in such systems. These two instabilities occur for the opposite directions of fluid flow. While the Dyakonov-Shur instability is characterized by the plasma frequency in 3D and the system size in 2D, the frequency of the entropy wave instability is tunable by the system size and the flow velocity.
\end{abstract}

\maketitle

{\it Introduction.---} Plasma instabilities attract significant attention and play an important role in various branches of science including high-energy and condensed matter physics, astrophysics, controlled thermonuclear fusion, etc. A few decades ago Dyakonov and Shur predicted~\cite{Dyakonov-Shur:1993} that an electron plasma in a hydrodynamic regime should become unstable in a biased two-dimensional (2D) heterostructure subject to a background direct current (dc) flow and rather unconventional asymmetric alternating current (ac) boundary conditions. This Dyakonov-Shur instability (DSI) appears due to the amplification of plasma waves (equivalently, plasmons) caused by multiple reflections from the device boundaries. Such an enhancement is reminiscent of the Fermi acceleration mechanism~\cite{Krymskii:1977,Bell:1978}, where charged particles are accelerated due to reflection from shock fronts or moving magnetic mirrors. The DSI could provide an effective way to create sources of terahertz radiation by using a direct current. This is particularly important in the modern industry where compact, efficient, and tunable sources of terahertz radiation are needed~\cite{Dhillon-Johnston:rev-2017}. Furthermore, the DSI allows one to detect terahertz radiation by converting an ac signal to the dc one~\cite{Dyakonov-Shur:1996}, which could be used to create terahertz detectors.

The recent surge of interest in the DSI is connected with the experimental observation of the electron hydrodynamics in 2D electron gas of $\mathrm{(Al,Ga)As}$ heterostructures~\cite{Molenkamp-Jong:1994,Jong-Molenkamp:1995} and graphene~\cite{Crossno-Fong:2016,Ghahari-Kim:2016,Krishna-Falkovich:2017,Berdyugin-Bandurin:2018,Bandurin-Falkovich:2018,Ku-Walsworth:2019,Sulpizio-Ilani:2019} (see Refs.~\cite{Lucas-Fong:rev-2017,Narozhny:rev-2019} for recent reviews on electron hydrodynamics). In addition to 2D systems, evidence of three-dimensional (3D) relativisticlike hydrodynamic electron transport was reported in the Weyl semimetal tungsten diphosphide WP$_2$~\cite{Gooth-Felser:2018}. Because Dirac and Weyl semimetals provide a suitable platform for investigating electron hydrodynamics in solids, the DSI in graphene received a lot of attention~\cite{Tomadin-Polini:2013,Svintsov-Otsuji:2013,Koseki-Satou:2016,Mendl-Polini:2019,Crabb-Aizin:2021}. Despite extensive theoretical studies, the generation of terahertz waves by the DSI was not confirmed experimentally yet. Nevertheless, the inverse effect, namely the rectification of ac signals, was reported in Refs.~\cite{Tauk-Shur:2006,Vitiello-Tredicucci:2012,Vicarelli-Polini-Tredicucci:2012,Giliberti-Evangelisti:2015,Bandurin-Svintsov:2018}.

Motivated by the recent experimental progress in Dirac and Weyl semimetals, we study hydrodynamic instabilities driven by a dc current in relativisticlike 2D and 3D systems with the Dyakonov-Shur boundary conditions amended by a fixed temperature boundary condition. Since the energy flow can be as important as the charge flow in a relativisticlike system, we pay special attention to the energy current in the hydrodynamic description. One of our main findings is an instability associated with the entropy waves~\cite{Landau:t6-2013,Ogilvie:2016}. We dub it the entropy wave instability (EWI). Unlike the conventional DSI, the frequency of the EWI is determined by the flow velocity. This makes the corresponding unstable modes easily tunable. Interestingly, the EWI and DSI occur for the currents of opposite directions.

{\it Model.---} In the hydrodynamic regime, the dynamics of the electron fluid made of relativisticlike quasiparticles is described by the Navier-Stokes equation, the charge and energy continuity relations, and the Gauss law that relates the electric potential to the charge density. The corresponding system of equations reads~\cite{Lucas-Fong:rev-2017,Narozhny:rev-2019}
\begin{eqnarray}
\label{DS-model-Eq1}
&&\frac{1}{v_F^2}\left[\partial_t +(\mathbf{u}\cdot\bm{\nabla})\right]\left(\mathbf{u} w\right)
+\frac{1}{v_F^2} w \mathbf{u}(\bm{\nabla}\cdot\mathbf{u})\nonumber\\
&&=-\bm{\nabla}P +en \bm{\nabla}\varphi +\eta \Delta \mathbf{u} +\eta \frac{d-2}{d} \bm{\nabla}\left(\bm{\nabla}\cdot\mathbf{u}\right)- \frac{w \mathbf{u}}{v_F^2\tau},\nonumber\\
\\
\label{DS-model-Eq2}
&&-e\partial_tn +\left(\bm{\nabla}\cdot\mathbf{J}\right)=0,\\
\label{DS-model-Eq3}
&&\partial_t\epsilon +\left(\bm{\nabla}\cdot\mathbf{J}^{\epsilon}\right)=\left(\mathbf{E}\cdot\mathbf{J}\right),\\
\label{DS-model-Eq4}
&&\Delta \varphi = 4\pi e \left(n-n_0\right).
\end{eqnarray}
Here, $w=\epsilon+P$ is the enthalpy, $\epsilon$ is the energy density, $P$ is the pressure, $\mathbf{u}$ is the electron fluid velocity, $n$ is the electron number density, $-e$ is the electron charge, and $v_F$ is the Fermi velocity. Notice that, because of a relativisticlike dispersion of quasiparticles, the thermodynamic quantities depend on the fluid velocity in the laboratory (ion lattice) frame. Their explicit expressions are given in the Supplemental Material~\cite{SM}. Unlike other quantities, the equilibrium charge densities in the laboratory and comoving frames are the same (i.e., $n=n_0$), since they must be compensated by the charge density of ions. From general considerations, the hydrodynamic regime is expected to break down when the fluid velocity approaches $v_F$. Therefore, we will assume that $u\ll v_F$.

In Eq.~(\ref{DS-model-Eq1}), the shear viscosity $\eta$ is defined as \mbox{$\eta = \eta_{\rm kin}w/v_F^2$}, where $\eta_{\rm kin}$ is the kinematic shear viscosity~\footnote{The bulk viscosity is vanishingly small in systems with relativisticlike quasiparticles, see, e.g., Ref.~\cite{Principi-Polini:2015} for an explicit calculation in graphene.}. The momentum relaxation is quantified by the relaxation time $\tau$ that describes scattering on impurities and phonons. In the hydrodynamic regime, we neglect the intrinsic electric and thermal conductivities~\cite{Lucas-Fong:rev-2017}. Therefore, the electric and energy current densities are proportional to fluid velocity $\mathbf{u}$, i.e.,  $\mathbf{J} = -en\mathbf{u}$ and $\mathbf{J}^{\epsilon}=w\mathbf{u}$.

To study current-driven instabilities, we employ the conventional linear stability analysis where weak fluctuations are superimposed on top of a steady uniform flow, quantified by the fluid velocity $\mathbf{u}_0 =u_0\hat{\mathbf{x}}$, e.g.,
\begin{equation}
\label{DS-model-dn}
u_x(t,\mathbf{r}) =u_0+ u_1 e^{-i\omega t+i\mathbf{k}\cdot\mathbf{r}}.
\end{equation}
Similar expressions are valid also for the other quantities ($n$, $\varphi$, and $\epsilon$). Here, $\omega$ and $\mathbf{k}$ are the angular frequency and the wave vector of excitations, respectively. For the sake of simplicity, we neglect all transverse fluctuations and focus on the one-dimensional instability assuming $\mathbf{k}=k_x \hat{\mathbf{x}}$. Linearizing Eqs.~(\ref{DS-model-Eq1})--(\ref{DS-model-Eq4}) and using ansatz (\ref{DS-model-dn}), we obtain the following set of linear algebraic equations at $\eta=0$:
\begin{widetext}
\begin{eqnarray}
\label{DS-model-eq-lin-1}
&&\left(\omega -2k_x u_0\right) u_1 +\frac{u_0}{w_0} \left(\omega -k_x u_0\right) w_1 -k_x \frac{v_F^2}{w_0} P_1 +i\frac{u_0}{\tau}\left(\frac{u_1}{u_0} +\frac{w_1}{w_0}-\frac{n_1}{n_0} \right) =-k_x\frac{e n_0 v_F^2}{w_0} \varphi_1,\\
\label{DS-model-eq-lin-2}
&&\left(\omega -k_x u_0\right) n_1 = k_x n_0 u_1,\\
\label{DS-model-eq-lin-3}
&&\omega \epsilon_1 -k_x u_0 w_1 -k_xw_0 u_1 -i\frac{1}{\tau}\frac{u_0^2}{v_F^2} w_0 \left(\frac{n_1}{n_0} +\frac{u_1}{u_0}\right) =-u_0 k_x en_0 \varphi_1,\\
\label{DS-model-eq-lin-4}
&&\mbox{3D:}\quad \varphi_1 = -\frac{4\pi e}{k^2} n_1 \quad \mbox{or} \quad \mbox{2D:}\quad \varphi_1 = -\frac{e}{C} n_1.
\end{eqnarray}
\end{widetext}
In the 2D system, we use the ``gradual channel" approximation~\cite{Shur:book,Dyakonov-Shur:1993}, where $C=\varepsilon/(4\pi L_g)$ is the capacitance per unit area, $\varepsilon$ is the dielectric constant of the substrate, and $L_g$ is the distance to the gate. Note that the steady-state solution of the hydrodynamic equations implies the presence of electric field $E_0 =- w_0 u_0/(en_0 v_F^2 \tau)$, which was explicitly used in Eqs.~(\ref{DS-model-eq-lin-1}) and (\ref{DS-model-eq-lin-3}).

In the dissipationless limit ($\tau\to\infty$) and to the leading order in $|u_0|/v_F$, the characteristic equation that determines the spectrum of collective modes reads (see Supplemental Material~\cite{SM} for details)
\begin{equation}
\label{DS-model-char-eq-3D}
\left(\omega -u_0k_x\right)\left[\omega^2 -\omega_p^2 +u_0 k_x\left(u_0k_x -\frac{4}{3}\omega\right)-v_s^2k_x^2\right]=0
\end{equation}
in 3D and
\begin{equation}
\label{DS-model-char-eq-2D}
\left(\omega -u_0k_x\right)\left[\omega\left(\omega - u_0 k_x\right) -v_s^2 k_x^2\left(1+ \xi\right) \right]=0
\end{equation}
in 2D. Here, $v_s = v_F/\sqrt{d}$ is the sound velocity in a $d$-dimensional space ($d=2,3$). The square of the plasma frequency for a 3D relativisticlike fluid is given by
\begin{equation}
\label{DS-model-omegap-def}
\omega_p^2 = \frac{4\pi e^2 n_0^2 v_F^2}{w_0}.
\end{equation}
In the 2D case, we introduced dimensionless parameter $\xi =2e^2 n_0^2/(w_0 C)$.

{\it Collective modes.---} To clarify the physical origin of the EWI, it is instructive to determine the solutions to the characteristic equations (\ref{DS-model-char-eq-3D}) and (\ref{DS-model-char-eq-2D}) in an infinite medium without imposing any boundary conditions.

To the leading order in $|u_0|/v_F\ll 1$, we find the following dispersion relations for collective modes:
\begin{eqnarray}
\label{DS-inf-omega-pm-3D}
\mbox{3D:}\quad \omega_{\pm} &\approx& \pm \sqrt{\omega_p^2 +v_s^2k_x^2} +\frac{2}{3} u_0 k_x,\\
\label{DS-inf-omega-pm-2D}
\mbox{2D:} \quad \omega_{\pm} &\approx& \pm v_p k_x +\frac{1}{2} u_0k_x,\\
\label{DS-inf-omega-e}
\mbox{2D and 3D:} \quad \omega_{e} &\approx& u_0 k_x.
\end{eqnarray}
Here, $\omega_{\pm}$ correspond to plasmons. Notice that the plasmon spectrum in a gated 2D sample is gapless and linear in the wave vector. The corresponding plasmon velocity is $v_p=v_s\sqrt{1+\xi}$. The third solution $\omega_{e}$ given in Eq.~(\ref{DS-inf-omega-e}) corresponds to the so-called entropy wave~\cite{Landau:t6-2013,Ogilvie:2016}.

Considering the important role of the entropy wave, let us discuss its properties in detail. For simplicity, we consider the limit $\tau \to \infty$. In drastic contrast to plasmons, the flow velocity does not oscillate in this wave, i.e., $u_1= 0$. It is characterized by oscillating electron number $n_1$ and energy $\epsilon_1$ densities and, in turn, the entropy. As follows from the Navier-Stokes equation (\ref{DS-model-eq-lin-1}), the solution with $u_1=0$ is possible because the gradient of pressure defined by the third term on the left-hand side of the equation is counterbalanced by the Coulomb force provided by the term on the right-hand side. To the leading order in $|u_0|/v_s$, one finds that
\begin{equation}
\label{DS-inf-sol-n1}
\frac{n_1}{n_0} \approx - \left(\frac{v_sk_x}{\omega_p}\right)^2 \frac{\epsilon_1}{w_0}.
\end{equation}
According to Eq.~(\ref{DS-inf-omega-e}), the entropy wave is a downstream wave; i.e., it propagates with the local flow velocity $u_0$.

It is interesting to point out that modes with similar dispersion relations appear in geophysics. For example, the Coriolis force and the pressure gradient compensate each other in the Rossby wave~\cite{Rossby:1938}, which is an inertial wave occurring in rotating fluids. Its dispersion relation contains the term $\propto u_0k_x$ and a term quadratic in $k_x$ due to the coordinate dependence of the planetary vorticity. In the absence of background flow, the frequency of the entropy mode is zero which is analogous to geostrophic currents~\cite{Gill:book}.

{\it Boundary conditions and instabilities.---} Both plasmons and entropy waves are stable collective modes in an infinite medium. Let us show now that the stability of these modes is affected profoundly by the boundary conditions. We consider a sample with length $L$ along the $x$ direction. In addition to the standard Dyakonov-Shur boundary conditions, we fix temperature at the left ($x=0$) surface
\begin{eqnarray}
\label{DS-inst-BC-n}
n_1(x=0) &=& 0,\\
\label{DS-inst-BC-J}
J_x(x=L) &\equiv& n_0u_1(x=L) +u_0n_1(x=L)= 0,\\
\label{DS-inst-BC-T}
T_1(x=0) &=& 0.
\end{eqnarray}
Physically the Dyakonov-Shur boundary conditions correspond to short circuiting the sample at $x=0$ (zero impedance)~(\ref{DS-inst-BC-n}) and leaving the other side $x=L$ open (infinite impedance)~(\ref{DS-inst-BC-J}). The condition in Eq.~(\ref{DS-inst-BC-T}) can be enforced by connecting the boundary to a large thermostat, e.g., made of a metal with high thermal conductivity and specific heat. 

It is convenient to solve hydrodynamic equations in terms of $u_1$, $n_1$, and $\epsilon_1$. One can show (see Supplemental Material~\cite{SM} for details) that oscillations of pressure $P_1$  are related to $\epsilon_1$ as follows:
\begin{equation}
\label{DS-inst-P1u-relat}
P_1 \approx \frac{\epsilon_1}{d} - \epsilon_0 \frac{2(d+1)}{d^3} \frac{u_1 u_0}{v_s^2},
\end{equation}
at the leading order in $|u_0|/v_s$. Then, the boundary condition (\ref{DS-inst-BC-T}) can be reexpressed in terms of $\epsilon_1$ and $u_1$, i.e.,
\begin{equation}
\label{DS-inst-BC-eps}
\epsilon_1(x=0) \approx \frac{u_1u_0}{v_s^2} \frac{d+1}{d^2} \epsilon_0 \left[1-(d+1)\left(1-\Lambda_p^2\right)\right],
\end{equation}
where $\Lambda_p = \omega_p/(v_s q_{\rm TF})$ and $q_{\rm TF}^2 = 4\pi e^2 \left(\partial_{\mu} n_0\right)$ is the square of the Thomas-Fermi wave vector.

We seek solutions to Eqs.~(\ref{DS-model-eq-lin-1})--(\ref{DS-model-eq-lin-4}) in the form
\begin{equation}
\label{DS-inst-sol-n}
\frac{n_1}{n_0} = \sum^3_{j=1}C_{j} e^{ik_{j} x}
\end{equation}
and define $u_1$, $\epsilon_1$, and $\varphi_1$ from Eqs.~(\ref{DS-model-eq-lin-2})--(\ref{DS-model-eq-lin-4}) (see Supplemental Material~\cite{SM} for the corresponding expressions). Here, $\sum_{j}$ runs over the three roots $k_j(\omega) $ of Eq.~(\ref{DS-model-char-eq-3D}) or (\ref{DS-model-char-eq-2D}). By using the boundary conditions (\ref{DS-inst-BC-n}), (\ref{DS-inst-BC-J}), and (\ref{DS-inst-BC-eps}), we derive the characteristic equation for $\omega$, which defines allowed collective modes in the system.

Let us start with the plasmons. To the linear order in $u_0$, their frequencies are given by the following relations:
\begin{equation}
\label{DS-inst-3D-omega-plasmon-app}
\omega^{3D}_{\pm} \approx
\pm \sqrt{\omega_p^2 +\left[v_s\frac{\pi}{L} \left(l+\frac{1}{2}\right)\right]^2} + i\frac{2u_0}{3L} \left(3-2\Lambda_p^2\right)
\end{equation}
in the 3D case, and
\begin{eqnarray}
\label{DS-inst-2D-omega-plasmon-app}
\omega^{2D}_{\pm} \approx \pm v_p\frac{\pi}{L}\left(l+\frac{1}{2}\right) +i\frac{u_0}{2L} \left(4-3\Lambda_p^2\right)
\end{eqnarray}
in the 2D case, respectively. In both expressions, $l\in \mathds{Z}$. (For the results to the quadratic order in $u_0$, see SM.) Since $0<\Lambda_p<1$ for $T\neq 0$ (see SM for the temperature dependence of $\Lambda_p$), the plasmons are unstable in both 3D and 2D systems. This is in agreement with Refs.~\cite{Svintsov-Otsuji:2013,Crabb-Aizin:2021} for slow flow
~\footnote{The growth rate for plasmons (\ref{DS-inf-omega-pm-2D}) differs from that in Ref.~\cite{Tomadin-Polini:2013}. The discrepancy can be traced to a different form of the nonlinear terms in the Navier-Stokes equation and the dependence of pressure $P$ on velocity.}.

As we see from Eqs.~(\ref{DS-inst-3D-omega-plasmon-app}) and (\ref{DS-inst-2D-omega-plasmon-app}), enforcing the boundary conditions leads to the DSI for $u_0>0$. In the linear regime, it is quantified by a growing amplitude $\propto e^{\mathrm{Im}\left[\omega_{\pm}\right] t}$. Eventually, the growth will be cut off by nonlinearities (see, e.g. Refs.~\cite{Mendl-Polini:2019,Crabb-Aizin:2021}). As for the real part, the plasmon frequencies are quantized due to the finite thickness of the slab, where $k_x\to\pi (l+1/2)/L$. As expected, the minimal frequency is determined by $\omega_p$ in 3D and the inverse sample size in 2D.

{\it Entropy wave instability and numerical results.---}
Let us turn to the entropy mode now. By solving the characteristic equation for large $L \omega_p/v_s$~\footnote{The hydrodynamic description is invalid for $L \omega_p/v_s\lesssim1$ because the system size becomes comparable to the inter-carrier distance in this case.}, the corresponding frequency can be approximated as
\begin{eqnarray}
\label{DS-inst-3D-omega-EW-app}
\omega^{3D}_{e} \approx \frac{2\pi l}{L} u_0 -i\frac{u_0 \omega_p}{v_s} -i \frac{u_0}{L} \ln{\left[\frac{3}{8} \frac{v_s^2}{u_0^2 \left(1-\Lambda_p^2\right)}\right]}
\end{eqnarray}
in 3D and
\begin{eqnarray}
\label{DS-inst-2D-omega-EW-app}
\omega^{2D}_{e} \approx \frac{2\pi l}{L} u_0 -i \frac{u_0}{L} \ln{\left[\frac{2}{3} \frac{v_p^2}{u_0^2 \left(1-\Lambda_p^2\right)}\right]}
\end{eqnarray}
in 2D. For the entropy wave, unlike plasmons, the real part of $\omega_e$  is controlled by the flow velocity and the sample size, i.e., $\mbox{Re}\left[\omega_{e}\right]\propto u_0/L$, in both 2D and 3D. The entropy mode becomes unstable for $u_0<0$ due to the combined effect of the fluid flow and the boundary conditions.

Let us emphasize several distinctions between the plasmon and entropy modes. Plasmons are characterized by large in-phase oscillations of energy and number densities, as well as having a non-negligible velocity (see Supplemental Material~\cite{SM} for details). They are also delocalized; i.e., the magnitude of oscillations is large throughout the slab. In the case of entropy modes, oscillations of velocity are suppressed. Unlike plasmons, these modes show a noticeable localization at the $x=L$ interface. This localization becomes less pronounced for the modes with large $l$ when the real part $|\mbox{Re}\left[\omega_{e}\right]|\gtrsim \omega_p$ in 3D or $|\mbox{Re}\left[\omega_{e}\right]|\gtrsim v_s\pi/(2L)$ in 2D, when the entropy waves may hybridize with plasmons.

\begin{figure}[ht!]
\centering
\subfigure{\includegraphics[width=0.45\textwidth]{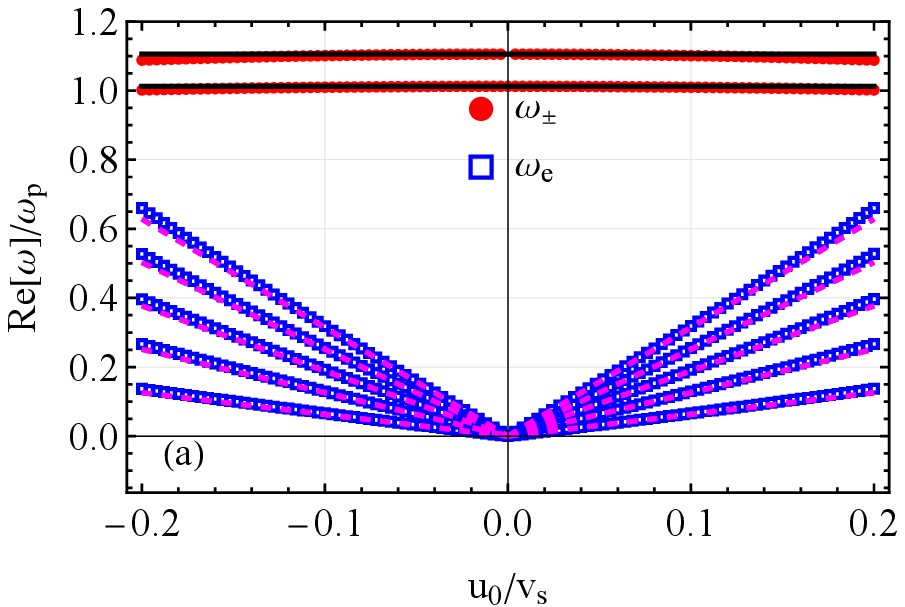}}
\hspace{0.05\textwidth}
\subfigure{\includegraphics[width=0.45\textwidth]{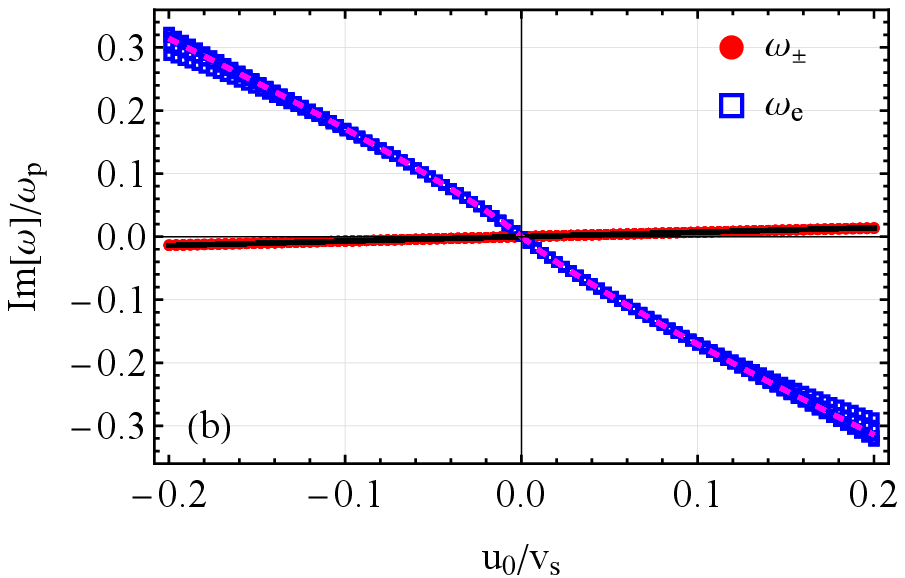}}
\caption{
The real (a) and imaginary (b) parts of the frequency of collective modes as a function of velocity $u_0$ in the 3D case. We show only the lowest five branches of numerical and approximate analytical results. Black solid lines correspond to the approximate expression (\ref{DS-inst-3D-omega-plasmon-app}). The approximate relation (\ref{DS-inst-3D-omega-EW-app}) is shown by magenta dashed lines. We fixed $L=10\, v_s/\omega_p$, $\Lambda_p \approx 0.98$, and took the limit $\tau\to\infty$.
}
\label{fig:DS-inst-3D}
\end{figure}

Our numerical and approximate analytical results for collective modes in a 3D Dirac system are shown in Fig.~\ref{fig:DS-inst-3D}. The results for the 2D case are qualitatively similar with the frequency scale normalized by $v_s/L$ instead of $\omega_p$ (see Supplemental Material~\cite{SM}). The separation between the branches of both modes become small for realistic system size $L \omega_p/v_s \gg1$. While this complicates the numerical calculations and obscures the presentation, the qualitative features remain the same as for $L \omega_p/v_s \sim 10$. Therefore, for the sake of clarity, we use a rather small width $L=10\, v_s/\omega_p$ and show only the lowest five branches of the numerical results and the approximate analytical solutions given in Eq.~(\ref{DS-inst-3D-omega-EW-app}). As one can see, even at such a small width, the density of solutions quickly increases at $u_0\to0$ for the entropy mode. It is clear from Fig.~\ref{fig:DS-inst-3D} that the approximate expressions (\ref{DS-inst-3D-omega-plasmon-app}) and  (\ref{DS-inst-3D-omega-EW-app}) agree well with the numerical results. As expected, the plasmons have nonzero frequencies at $u_0\to0$ and the solutions for the entropy modes vanish in this limit. We notice also that the instability increment is much larger for the entropy waves than for plasmons. Therefore, the corresponding instability should be more pronounced than the DSI for the same flow velocities.

All in all, different frequencies, growth rates, and spatial profiles of oscillating variables make the EWI profoundly different from the conventional DSI.

{\it Estimates and momentum relaxation effects.---}
For typical 3D Dirac and Weyl semimetal parameters, we use $\mu_0=20~\mbox{meV}$, $T_0=25~\mbox{K}$, and the Fermi velocity $v_F\approx1.4\times10^7~\mbox{cm/s}$~\cite{Kumar-Felser:2017}. Then, we estimate $\omega_p/(2\pi)\approx 12~\mbox{THz}$ and $v_s\approx8\times10^{6}~\mbox{cm/s}$. The characteristic length scale is $v_s/\omega_p \approx 1~\mbox{nm}$.

For the 2D case, we use graphene as a characteristic system with $v_F=1.1\times10^{8}~\mbox{cm/s}$, $\mu_0=100~\mbox{meV}$, $T_0=100~\mbox{K}$, $L_{g}=100~\mbox{nm}$, and $\varepsilon=3.3$ (assuming a hexagonal boron nitride substrate). In this case, $v_s\approx7.8\times10^{7}~\mbox{cm/s}$, $\xi\approx65.1$, and $v_{p}\approx 8.1 \, v_s$. The corresponding characteristic frequency of collective modes is $v_p/L \approx \left(1~\mu\mbox{m}/L\right)~\mbox{THz}$, i.e., it also lies in the terahertz range.

It is instructive to discuss briefly the effects of momentum relaxation and viscosity. In view of a distinct nature of plasmons and entropy waves, the role of momentum relaxation in the DSI and EWI is qualitatively different. The suppression of the plasmon DSI can be roughly described by replacing $\omega\to\omega-i/\tau$, where $\tau$ is the relaxation time. For the parameters used, the instability disappears when $\tau\lesssim 3 L/(2u_0)$ in 3D and $\tau\lesssim 2L/u_0$ in 2D (we assumed $\Lambda_p\approx1$ here). On the other hand, the EWI is quite robust with respect to momentum relaxation. This is explained by the fact that these waves have weak oscillating velocity compared to other oscillating variables, e.g., $|u_1|/v_s \ll |n_1|/n_0$; see also Supplemental Material. The effects of viscosity can be estimated similar to Ref.~\cite{Dyakonov-Shur:1993} as $\omega-i/\tau\to\omega-i/\tau -i \eta_{\rm kin} \pi^2/L^2$. In essence, it also suppresses the DSI but becomes important only for a small width. A more detailed study of the dissipation effects will be reported elsewhere.

{\it Summary.---}
We found that Dirac and Weyl semimetals, subject to the Dyakonov-Shur boundary conditions and a boundary condition for temperature, develop an entropy wave instability. The latter is connected with the entropy mode in relativisticlike hydrodynamics, where the energy current plays an important role. The entropy wave instability is absent in materials with a nonrelativistic energy dispersion, where the energy current plays a secondary role.

We estimate that the growth rate of the entropy wave instability is parametrically larger than that for the Dyakonov-Shur instability. Moreover, the two instabilities occur for the opposite directions of the applied current. The frequencies of unstable modes are determined by the system size and the flow velocity (entropy wave), only the system size (2D plasmons), and the plasmon frequency (3D plasmons). The tunability of the entropy wave instability provides the means to detect and distinguish it from other instabilities in the experiment. It can be identified by measuring the emission of radiation with a frequency proportional to the flow velocity. Our estimates suggest that the current-driven instabilities are achievable for realistic samples and flow velocities. Thus, the entropy wave instability holds a potential for use in tunable sources of radiation.

\begin{acknowledgments}
We are grateful to Dmitry~Svintsov for bringing our attention to the problem of the DSI in 3D systems and for useful discussions. P.O.S. acknowledges the support through the Yale Prize Postdoctoral Fellowship in Condensed Matter Theory. The work of I.A.S. was supported by the U.S. National Science Foundation under Grant No.~PHY-1713950.
\end{acknowledgments}

\bibliography{library-short}

\begin{thebibliography}{41}%
\makeatletter
\providecommand \@ifxundefined [1]{%
 \@ifx{#1\undefined}
}%
\providecommand \@ifnum [1]{%
 \ifnum #1\expandafter \@firstoftwo
 \else \expandafter \@secondoftwo
 \fi
}%
\providecommand \@ifx [1]{%
 \ifx #1\expandafter \@firstoftwo
 \else \expandafter \@secondoftwo
 \fi
}%
\providecommand \natexlab [1]{#1}%
\providecommand \enquote  [1]{``#1''}%
\providecommand \bibnamefont  [1]{#1}%
\providecommand \bibfnamefont [1]{#1}%
\providecommand \citenamefont [1]{#1}%
\providecommand \href@noop [0]{\@secondoftwo}%
\providecommand \href [0]{\begingroup \@sanitize@url \@href}%
\providecommand \@href[1]{\@@startlink{#1}\@@href}%
\providecommand \@@href[1]{\endgroup#1\@@endlink}%
\providecommand \@sanitize@url [0]{\catcode `\\12\catcode `\$12\catcode
  `\&12\catcode `\#12\catcode `\^12\catcode `\_12\catcode `\%12\relax}%
\providecommand \@@startlink[1]{}%
\providecommand \@@endlink[0]{}%
\providecommand \url  [0]{\begingroup\@sanitize@url \@url }%
\providecommand \@url [1]{\endgroup\@href {#1}{\urlprefix }}%
\providecommand \urlprefix  [0]{URL }%
\providecommand \Eprint [0]{\href }%
\providecommand \doibase [0]{https://doi.org/}%
\providecommand \selectlanguage [0]{\@gobble}%
\providecommand \bibinfo  [0]{\@secondoftwo}%
\providecommand \bibfield  [0]{\@secondoftwo}%
\providecommand \translation [1]{[#1]}%
\providecommand \BibitemOpen [0]{}%
\providecommand \bibitemStop [0]{}%
\providecommand \bibitemNoStop [0]{.\EOS\space}%
\providecommand \EOS [0]{\spacefactor3000\relax}%
\providecommand \BibitemShut  [1]{\csname bibitem#1\endcsname}%
\let\auto@bib@innerbib\@empty
\bibitem [{\citenamefont {Dyakonov}\ and\ \citenamefont
  {Shur}(1993)}]{Dyakonov-Shur:1993}%
  \BibitemOpen
  \bibfield  {author} {\bibinfo {author} {\bibfnamefont {M.}~\bibnamefont
  {Dyakonov}}\ and\ \bibinfo {author} {\bibfnamefont {M.}~\bibnamefont
  {Shur}},\ }\bibfield  {title} {\bibinfo {title} {{Shallow water analogy for a
  ballistic field effect transistor: New mechanism of plasma wave generation by
  dc current}},\ }\href {https://doi.org/10.1103/PhysRevLett.71.2465}
  {\bibfield  {journal} {\bibinfo  {journal} {Phys. Rev. Lett.}\ }\textbf
  {\bibinfo {volume} {71}},\ \bibinfo {pages} {2465} (\bibinfo {year}
  {1993})}\BibitemShut {NoStop}%
\bibitem [{\citenamefont {Krymskii}(1977)}]{Krymskii:1977}%
  \BibitemOpen
  \bibfield  {author} {\bibinfo {author} {\bibfnamefont {G.~F.}\ \bibnamefont
  {Krymskii}},\ }\bibfield  {title} {\bibinfo {title} {{A regular mechanism for
  the acceleration of charged particles on the front of a shock wave}},\
  }\href@noop {} {\bibfield  {journal} {\bibinfo  {journal} {Sov. Phys. -
  Dokl.}\ }\textbf {\bibinfo {volume} {22}},\ \bibinfo {pages} {327} (\bibinfo
  {year} {1977})}\BibitemShut {NoStop}%
\bibitem [{\citenamefont {Bell}(1978)}]{Bell:1978}%
  \BibitemOpen
  \bibfield  {author} {\bibinfo {author} {\bibfnamefont {A.~R.}\ \bibnamefont
  {Bell}},\ }\bibfield  {title} {\bibinfo {title} {{The acceleration of cosmic
  rays in shock fronts – I}},\ }\href
  {https://doi.org/10.1093/mnras/182.2.147} {\bibfield  {journal} {\bibinfo
  {journal} {Mon. Not. R. Astron. Soc.}\ }\textbf {\bibinfo {volume} {182}},\
  \bibinfo {pages} {147} (\bibinfo {year} {1978})}\BibitemShut {NoStop}%
\bibitem [{\citenamefont {Dhillon}\ \emph {et~al.}(2017)\citenamefont
  {Dhillon}, \citenamefont {Vitiello}, \citenamefont {Linfield}, \citenamefont
  {Davies}, \citenamefont {Hoffmann}, \citenamefont {Booske}, \citenamefont
  {Paoloni}, \citenamefont {Gensch}, \citenamefont {Weightman}, \citenamefont
  {Williams}, \citenamefont {Castro-Camus}, \citenamefont {Cumming},
  \citenamefont {Simoens}, \citenamefont {Escorcia-Carranza}, \citenamefont
  {Grant}, \citenamefont {Lucyszyn}, \citenamefont {Kuwata-Gonokami},
  \citenamefont {Konishi}, \citenamefont {Koch}, \citenamefont {Schmuttenmaer},
  \citenamefont {Cocker}, \citenamefont {Huber}, \citenamefont {Markelz},
  \citenamefont {Taylor}, \citenamefont {Wallace}, \citenamefont {{Axel
  Zeitler}}, \citenamefont {Sibik}, \citenamefont {Korter}, \citenamefont
  {Ellison}, \citenamefont {Rea}, \citenamefont {Goldsmith}, \citenamefont
  {Cooper}, \citenamefont {Appleby}, \citenamefont {Pardo}, \citenamefont
  {Huggard}, \citenamefont {Krozer}, \citenamefont {Shams}, \citenamefont
  {Fice}, \citenamefont {Renaud}, \citenamefont {Seeds}, \citenamefont
  {St{\"{o}}hr}, \citenamefont {Naftaly}, \citenamefont {Ridler}, \citenamefont
  {Clarke}, \citenamefont {Cunningham},\ and\ \citenamefont
  {Johnston}}]{Dhillon-Johnston:rev-2017}%
  \BibitemOpen
  \bibfield  {author} {\bibinfo {author} {\bibfnamefont {S.~S.}\ \bibnamefont
  {Dhillon}}, \bibinfo {author} {\bibfnamefont {M.~S.}\ \bibnamefont
  {Vitiello}}, \bibinfo {author} {\bibfnamefont {E.~H.}\ \bibnamefont
  {Linfield}}, \bibinfo {author} {\bibfnamefont {A.~G.}\ \bibnamefont
  {Davies}}, \bibinfo {author} {\bibfnamefont {M.~C.}\ \bibnamefont
  {Hoffmann}}, \bibinfo {author} {\bibfnamefont {J.}~\bibnamefont {Booske}},
  \bibinfo {author} {\bibfnamefont {C.}~\bibnamefont {Paoloni}}, \bibinfo
  {author} {\bibfnamefont {M.}~\bibnamefont {Gensch}}, \bibinfo {author}
  {\bibfnamefont {P.}~\bibnamefont {Weightman}}, \bibinfo {author}
  {\bibfnamefont {G.~P.}\ \bibnamefont {Williams}}, \bibinfo {author}
  {\bibfnamefont {E.}~\bibnamefont {Castro-Camus}}, \bibinfo {author}
  {\bibfnamefont {D.~R.}\ \bibnamefont {Cumming}}, \bibinfo {author}
  {\bibfnamefont {F.}~\bibnamefont {Simoens}}, \bibinfo {author} {\bibfnamefont
  {I.}~\bibnamefont {Escorcia-Carranza}}, \bibinfo {author} {\bibfnamefont
  {J.}~\bibnamefont {Grant}}, \bibinfo {author} {\bibfnamefont
  {S.}~\bibnamefont {Lucyszyn}}, \bibinfo {author} {\bibfnamefont
  {M.}~\bibnamefont {Kuwata-Gonokami}}, \bibinfo {author} {\bibfnamefont
  {K.}~\bibnamefont {Konishi}}, \bibinfo {author} {\bibfnamefont
  {M.}~\bibnamefont {Koch}}, \bibinfo {author} {\bibfnamefont {C.~A.}\
  \bibnamefont {Schmuttenmaer}}, \bibinfo {author} {\bibfnamefont {T.~L.}\
  \bibnamefont {Cocker}}, \bibinfo {author} {\bibfnamefont {R.}~\bibnamefont
  {Huber}}, \bibinfo {author} {\bibfnamefont {A.~G.}\ \bibnamefont {Markelz}},
  \bibinfo {author} {\bibfnamefont {Z.~D.}\ \bibnamefont {Taylor}}, \bibinfo
  {author} {\bibfnamefont {V.~P.}\ \bibnamefont {Wallace}}, \bibinfo {author}
  {\bibfnamefont {J.}~\bibnamefont {{Axel Zeitler}}}, \bibinfo {author}
  {\bibfnamefont {J.}~\bibnamefont {Sibik}}, \bibinfo {author} {\bibfnamefont
  {T.~M.}\ \bibnamefont {Korter}}, \bibinfo {author} {\bibfnamefont
  {B.}~\bibnamefont {Ellison}}, \bibinfo {author} {\bibfnamefont
  {S.}~\bibnamefont {Rea}}, \bibinfo {author} {\bibfnamefont {P.}~\bibnamefont
  {Goldsmith}}, \bibinfo {author} {\bibfnamefont {K.~B.}\ \bibnamefont
  {Cooper}}, \bibinfo {author} {\bibfnamefont {R.}~\bibnamefont {Appleby}},
  \bibinfo {author} {\bibfnamefont {D.}~\bibnamefont {Pardo}}, \bibinfo
  {author} {\bibfnamefont {P.~G.}\ \bibnamefont {Huggard}}, \bibinfo {author}
  {\bibfnamefont {V.}~\bibnamefont {Krozer}}, \bibinfo {author} {\bibfnamefont
  {H.}~\bibnamefont {Shams}}, \bibinfo {author} {\bibfnamefont
  {M.}~\bibnamefont {Fice}}, \bibinfo {author} {\bibfnamefont {C.}~\bibnamefont
  {Renaud}}, \bibinfo {author} {\bibfnamefont {A.}~\bibnamefont {Seeds}},
  \bibinfo {author} {\bibfnamefont {A.}~\bibnamefont {St{\"{o}}hr}}, \bibinfo
  {author} {\bibfnamefont {M.}~\bibnamefont {Naftaly}}, \bibinfo {author}
  {\bibfnamefont {N.}~\bibnamefont {Ridler}}, \bibinfo {author} {\bibfnamefont
  {R.}~\bibnamefont {Clarke}}, \bibinfo {author} {\bibfnamefont {J.~E.}\
  \bibnamefont {Cunningham}},\ and\ \bibinfo {author} {\bibfnamefont {M.~B.}\
  \bibnamefont {Johnston}},\ }\bibfield  {title} {\bibinfo {title} {{The 2017
  terahertz science and technology roadmap}},\ }\href
  {https://doi.org/10.1088/1361-6463/50/4/043001} {\bibfield  {journal}
  {\bibinfo  {journal} {J. Phys. D. Appl. Phys.}\ }\textbf {\bibinfo {volume}
  {50}},\ \bibinfo {pages} {043001} (\bibinfo {year} {2017})}\BibitemShut
  {NoStop}%
\bibitem [{\citenamefont {Dyakonov}\ and\ \citenamefont
  {Shur}(1996)}]{Dyakonov-Shur:1996}%
  \BibitemOpen
  \bibfield  {author} {\bibinfo {author} {\bibfnamefont {M.}~\bibnamefont
  {Dyakonov}}\ and\ \bibinfo {author} {\bibfnamefont {M.}~\bibnamefont
  {Shur}},\ }\bibfield  {title} {\bibinfo {title} {{Detection, mixing, and
  frequency multiplication of terahertz radiation by two-dimensional electronic
  fluid}},\ }\href {https://doi.org/10.1109/16.485650} {\bibfield  {journal}
  {\bibinfo  {journal} {IEEE Trans. Electron Dev.}\ }\textbf {\bibinfo {volume}
  {43}},\ \bibinfo {pages} {380} (\bibinfo {year} {1996})}\BibitemShut
  {NoStop}%
\bibitem [{\citenamefont {Molenkamp}\ and\ \citenamefont
  {de~Jong}(1994)}]{Molenkamp-Jong:1994}%
  \BibitemOpen
  \bibfield  {author} {\bibinfo {author} {\bibfnamefont {L.~W.}\ \bibnamefont
  {Molenkamp}}\ and\ \bibinfo {author} {\bibfnamefont {M.~J.}\ \bibnamefont
  {de~Jong}},\ }\bibfield  {title} {\bibinfo {title} {{Observation of Knudsen
  and Gurzhi transport regimes in a two-dimensional wire}},\ }\href
  {https://doi.org/10.1016/0038-1101(94)90244-5} {\bibfield  {journal}
  {\bibinfo  {journal} {Solid State Electron.}\ }\textbf {\bibinfo {volume}
  {37}},\ \bibinfo {pages} {551} (\bibinfo {year} {1994})}\BibitemShut
  {NoStop}%
\bibitem [{\citenamefont {{De Jong}}\ and\ \citenamefont
  {Molenkamp}(1995)}]{Jong-Molenkamp:1995}%
  \BibitemOpen
  \bibfield  {author} {\bibinfo {author} {\bibfnamefont {M.~J.~M.}\
  \bibnamefont {{De Jong}}}\ and\ \bibinfo {author} {\bibfnamefont {L.~W.}\
  \bibnamefont {Molenkamp}},\ }\bibfield  {title} {\bibinfo {title}
  {{Hydrodynamic electron flow in high-mobility wires}},\ }\href
  {https://doi.org/10.1103/PhysRevB.51.13389} {\bibfield  {journal} {\bibinfo
  {journal} {Phys. Rev. B}\ }\textbf {\bibinfo {volume} {51}},\ \bibinfo
  {pages} {13389} (\bibinfo {year} {1995})}\BibitemShut {NoStop}%
\bibitem [{\citenamefont {Crossno}\ \emph {et~al.}(2016)\citenamefont
  {Crossno}, \citenamefont {Shi}, \citenamefont {Wang}, \citenamefont {Liu},
  \citenamefont {Harzheim}, \citenamefont {Lucas}, \citenamefont {Sachdev},
  \citenamefont {Kim}, \citenamefont {Taniguchi}, \citenamefont {Watanabe},
  \citenamefont {Ohki},\ and\ \citenamefont {Fong}}]{Crossno-Fong:2016}%
  \BibitemOpen
  \bibfield  {author} {\bibinfo {author} {\bibfnamefont {J.}~\bibnamefont
  {Crossno}}, \bibinfo {author} {\bibfnamefont {J.~K.}\ \bibnamefont {Shi}},
  \bibinfo {author} {\bibfnamefont {K.}~\bibnamefont {Wang}}, \bibinfo {author}
  {\bibfnamefont {X.}~\bibnamefont {Liu}}, \bibinfo {author} {\bibfnamefont
  {A.}~\bibnamefont {Harzheim}}, \bibinfo {author} {\bibfnamefont
  {A.}~\bibnamefont {Lucas}}, \bibinfo {author} {\bibfnamefont
  {S.}~\bibnamefont {Sachdev}}, \bibinfo {author} {\bibfnamefont
  {P.}~\bibnamefont {Kim}}, \bibinfo {author} {\bibfnamefont {T.}~\bibnamefont
  {Taniguchi}}, \bibinfo {author} {\bibfnamefont {K.}~\bibnamefont {Watanabe}},
  \bibinfo {author} {\bibfnamefont {T.~A.}\ \bibnamefont {Ohki}},\ and\
  \bibinfo {author} {\bibfnamefont {K.~C.}\ \bibnamefont {Fong}},\ }\bibfield
  {title} {\bibinfo {title} {{Observation of the Dirac fluid and the breakdown
  of the Wiedemann-Franz law in graphene}},\ }\href
  {https://doi.org/10.1126/science.aad0343} {\bibfield  {journal} {\bibinfo
  {journal} {Science}\ }\textbf {\bibinfo {volume} {351}},\ \bibinfo {pages}
  {1058} (\bibinfo {year} {2016})}\BibitemShut {NoStop}%
\bibitem [{\citenamefont {Ghahari}\ \emph {et~al.}(2016)\citenamefont
  {Ghahari}, \citenamefont {Xie}, \citenamefont {Taniguchi}, \citenamefont
  {Watanabe}, \citenamefont {Foster},\ and\ \citenamefont
  {Kim}}]{Ghahari-Kim:2016}%
  \BibitemOpen
  \bibfield  {author} {\bibinfo {author} {\bibfnamefont {F.}~\bibnamefont
  {Ghahari}}, \bibinfo {author} {\bibfnamefont {H.-Y.}\ \bibnamefont {Xie}},
  \bibinfo {author} {\bibfnamefont {T.}~\bibnamefont {Taniguchi}}, \bibinfo
  {author} {\bibfnamefont {K.}~\bibnamefont {Watanabe}}, \bibinfo {author}
  {\bibfnamefont {M.~S.}\ \bibnamefont {Foster}},\ and\ \bibinfo {author}
  {\bibfnamefont {P.}~\bibnamefont {Kim}},\ }\bibfield  {title} {\bibinfo
  {title} {{Enhanced Thermoelectric Power in Graphene: Violation of the Mott
  Relation by Inelastic Scattering}},\ }\href
  {https://doi.org/10.1103/PhysRevLett.116.136802} {\bibfield  {journal}
  {\bibinfo  {journal} {Phys. Rev. Lett.}\ }\textbf {\bibinfo {volume} {116}},\
  \bibinfo {pages} {136802} (\bibinfo {year} {2016})}\BibitemShut {NoStop}%
\bibitem [{\citenamefont {{Krishna Kumar}}\ \emph {et~al.}(2017)\citenamefont
  {{Krishna Kumar}}, \citenamefont {Bandurin}, \citenamefont {Pellegrino},
  \citenamefont {Cao}, \citenamefont {Principi}, \citenamefont {Guo},
  \citenamefont {Auton}, \citenamefont {{Ben Shalom}}, \citenamefont
  {Ponomarenko}, \citenamefont {Falkovich}, \citenamefont {Watanabe},
  \citenamefont {Taniguchi}, \citenamefont {Grigorieva}, \citenamefont
  {Levitov}, \citenamefont {Polini},\ and\ \citenamefont
  {Geim}}]{Krishna-Falkovich:2017}%
  \BibitemOpen
  \bibfield  {author} {\bibinfo {author} {\bibfnamefont {R.}~\bibnamefont
  {{Krishna Kumar}}}, \bibinfo {author} {\bibfnamefont {D.~A.}\ \bibnamefont
  {Bandurin}}, \bibinfo {author} {\bibfnamefont {F.~M.~D.}\ \bibnamefont
  {Pellegrino}}, \bibinfo {author} {\bibfnamefont {Y.}~\bibnamefont {Cao}},
  \bibinfo {author} {\bibfnamefont {A.}~\bibnamefont {Principi}}, \bibinfo
  {author} {\bibfnamefont {H.}~\bibnamefont {Guo}}, \bibinfo {author}
  {\bibfnamefont {G.~H.}\ \bibnamefont {Auton}}, \bibinfo {author}
  {\bibfnamefont {M.}~\bibnamefont {{Ben Shalom}}}, \bibinfo {author}
  {\bibfnamefont {L.~A.}\ \bibnamefont {Ponomarenko}}, \bibinfo {author}
  {\bibfnamefont {G.}~\bibnamefont {Falkovich}}, \bibinfo {author}
  {\bibfnamefont {K.}~\bibnamefont {Watanabe}}, \bibinfo {author}
  {\bibfnamefont {T.}~\bibnamefont {Taniguchi}}, \bibinfo {author}
  {\bibfnamefont {I.~V.}\ \bibnamefont {Grigorieva}}, \bibinfo {author}
  {\bibfnamefont {L.~S.}\ \bibnamefont {Levitov}}, \bibinfo {author}
  {\bibfnamefont {M.}~\bibnamefont {Polini}},\ and\ \bibinfo {author}
  {\bibfnamefont {A.~K.}\ \bibnamefont {Geim}},\ }\bibfield  {title} {\bibinfo
  {title} {{Superballistic flow of viscous electron fluid through graphene
  constrictions}},\ }\href {https://doi.org/10.1038/nphys4240} {\bibfield
  {journal} {\bibinfo  {journal} {Nat. Phys.}\ }\textbf {\bibinfo {volume}
  {13}},\ \bibinfo {pages} {1182} (\bibinfo {year} {2017})}\BibitemShut
  {NoStop}%
\bibitem [{\citenamefont {Berdyugin}\ \emph {et~al.}(2019)\citenamefont
  {Berdyugin}, \citenamefont {Xu}, \citenamefont {Pellegrino}, \citenamefont
  {{Krishna Kumar}}, \citenamefont {Principi}, \citenamefont {Torre},
  \citenamefont {{Ben Shalom}}, \citenamefont {Taniguchi}, \citenamefont
  {Watanabe}, \citenamefont {Grigorieva}, \citenamefont {Polini}, \citenamefont
  {Geim},\ and\ \citenamefont {Bandurin}}]{Berdyugin-Bandurin:2018}%
  \BibitemOpen
  \bibfield  {author} {\bibinfo {author} {\bibfnamefont {A.~I.}\ \bibnamefont
  {Berdyugin}}, \bibinfo {author} {\bibfnamefont {S.~G.}\ \bibnamefont {Xu}},
  \bibinfo {author} {\bibfnamefont {F.~M.~D.}\ \bibnamefont {Pellegrino}},
  \bibinfo {author} {\bibfnamefont {R.}~\bibnamefont {{Krishna Kumar}}},
  \bibinfo {author} {\bibfnamefont {A.}~\bibnamefont {Principi}}, \bibinfo
  {author} {\bibfnamefont {I.}~\bibnamefont {Torre}}, \bibinfo {author}
  {\bibfnamefont {M.}~\bibnamefont {{Ben Shalom}}}, \bibinfo {author}
  {\bibfnamefont {T.}~\bibnamefont {Taniguchi}}, \bibinfo {author}
  {\bibfnamefont {K.}~\bibnamefont {Watanabe}}, \bibinfo {author}
  {\bibfnamefont {I.~V.}\ \bibnamefont {Grigorieva}}, \bibinfo {author}
  {\bibfnamefont {M.}~\bibnamefont {Polini}}, \bibinfo {author} {\bibfnamefont
  {A.~K.}\ \bibnamefont {Geim}},\ and\ \bibinfo {author} {\bibfnamefont
  {D.~A.}\ \bibnamefont {Bandurin}},\ }\bibfield  {title} {\bibinfo {title}
  {{Measuring Hall viscosity of graphene's electron fluid}},\ }\href
  {https://doi.org/10.1126/science.aau0685} {\bibfield  {journal} {\bibinfo
  {journal} {Science}\ }\textbf {\bibinfo {volume} {364}},\ \bibinfo {pages}
  {162} (\bibinfo {year} {2019})}\BibitemShut {NoStop}%
\bibitem [{\citenamefont {Bandurin}\ \emph
  {et~al.}(2018{\natexlab{a}})\citenamefont {Bandurin}, \citenamefont {Shytov},
  \citenamefont {Levitov}, \citenamefont {Kumar}, \citenamefont {Berdyugin},
  \citenamefont {{Ben Shalom}}, \citenamefont {Grigorieva}, \citenamefont
  {Geim},\ and\ \citenamefont {Falkovich}}]{Bandurin-Falkovich:2018}%
  \BibitemOpen
  \bibfield  {author} {\bibinfo {author} {\bibfnamefont {D.~A.}\ \bibnamefont
  {Bandurin}}, \bibinfo {author} {\bibfnamefont {A.~V.}\ \bibnamefont
  {Shytov}}, \bibinfo {author} {\bibfnamefont {L.~S.}\ \bibnamefont {Levitov}},
  \bibinfo {author} {\bibfnamefont {R.~K.}\ \bibnamefont {Kumar}}, \bibinfo
  {author} {\bibfnamefont {A.~I.}\ \bibnamefont {Berdyugin}}, \bibinfo {author}
  {\bibfnamefont {M.}~\bibnamefont {{Ben Shalom}}}, \bibinfo {author}
  {\bibfnamefont {I.~V.}\ \bibnamefont {Grigorieva}}, \bibinfo {author}
  {\bibfnamefont {A.~K.}\ \bibnamefont {Geim}},\ and\ \bibinfo {author}
  {\bibfnamefont {G.}~\bibnamefont {Falkovich}},\ }\bibfield  {title} {\bibinfo
  {title} {{Fluidity onset in graphene}},\ }\href
  {https://doi.org/10.1038/s41467-018-07004-4} {\bibfield  {journal} {\bibinfo
  {journal} {Nat. Commun.}\ }\textbf {\bibinfo {volume} {9}},\ \bibinfo {pages}
  {4533} (\bibinfo {year} {2018}{\natexlab{a}})}\BibitemShut {NoStop}%
\bibitem [{\citenamefont {Ku}\ \emph {et~al.}(2020)\citenamefont {Ku},
  \citenamefont {Zhou}, \citenamefont {Li}, \citenamefont {Shin}, \citenamefont
  {Shi}, \citenamefont {Burch}, \citenamefont {Anderson}, \citenamefont
  {Pierce}, \citenamefont {Xie}, \citenamefont {Hamo}, \citenamefont {Vool},
  \citenamefont {Zhang}, \citenamefont {Casola}, \citenamefont {Taniguchi},
  \citenamefont {Watanabe}, \citenamefont {Fogler}, \citenamefont {Kim},
  \citenamefont {Yacoby},\ and\ \citenamefont {Walsworth}}]{Ku-Walsworth:2019}%
  \BibitemOpen
  \bibfield  {author} {\bibinfo {author} {\bibfnamefont {M.~J.~H.}\
  \bibnamefont {Ku}}, \bibinfo {author} {\bibfnamefont {T.~X.}\ \bibnamefont
  {Zhou}}, \bibinfo {author} {\bibfnamefont {Q.}~\bibnamefont {Li}}, \bibinfo
  {author} {\bibfnamefont {Y.~J.}\ \bibnamefont {Shin}}, \bibinfo {author}
  {\bibfnamefont {J.~K.}\ \bibnamefont {Shi}}, \bibinfo {author} {\bibfnamefont
  {C.}~\bibnamefont {Burch}}, \bibinfo {author} {\bibfnamefont {L.~E.}\
  \bibnamefont {Anderson}}, \bibinfo {author} {\bibfnamefont {A.~T.}\
  \bibnamefont {Pierce}}, \bibinfo {author} {\bibfnamefont {Y.}~\bibnamefont
  {Xie}}, \bibinfo {author} {\bibfnamefont {A.}~\bibnamefont {Hamo}}, \bibinfo
  {author} {\bibfnamefont {U.}~\bibnamefont {Vool}}, \bibinfo {author}
  {\bibfnamefont {H.}~\bibnamefont {Zhang}}, \bibinfo {author} {\bibfnamefont
  {F.}~\bibnamefont {Casola}}, \bibinfo {author} {\bibfnamefont
  {T.}~\bibnamefont {Taniguchi}}, \bibinfo {author} {\bibfnamefont
  {K.}~\bibnamefont {Watanabe}}, \bibinfo {author} {\bibfnamefont {M.~M.}\
  \bibnamefont {Fogler}}, \bibinfo {author} {\bibfnamefont {P.}~\bibnamefont
  {Kim}}, \bibinfo {author} {\bibfnamefont {A.}~\bibnamefont {Yacoby}},\ and\
  \bibinfo {author} {\bibfnamefont {R.~L.}\ \bibnamefont {Walsworth}},\
  }\bibfield  {title} {\bibinfo {title} {{Imaging viscous flow of the Dirac
  fluid in graphene}},\ }\href {https://doi.org/10.1038/s41586-020-2507-2}
  {\bibfield  {journal} {\bibinfo  {journal} {Nature (London)}\ }\textbf
  {\bibinfo {volume} {583}},\ \bibinfo {pages} {537} (\bibinfo {year}
  {2020})}\BibitemShut {NoStop}%
\bibitem [{\citenamefont {Sulpizio}\ \emph {et~al.}(2019)\citenamefont
  {Sulpizio}, \citenamefont {Ella}, \citenamefont {Rozen}, \citenamefont
  {Birkbeck}, \citenamefont {Perello}, \citenamefont {Dutta}, \citenamefont
  {Ben-Shalom}, \citenamefont {Taniguchi}, \citenamefont {Watanabe},
  \citenamefont {Holder}, \citenamefont {Queiroz}, \citenamefont {Principi},
  \citenamefont {Stern}, \citenamefont {Scaffidi}, \citenamefont {Geim},\ and\
  \citenamefont {Ilani}}]{Sulpizio-Ilani:2019}%
  \BibitemOpen
  \bibfield  {author} {\bibinfo {author} {\bibfnamefont {J.~A.}\ \bibnamefont
  {Sulpizio}}, \bibinfo {author} {\bibfnamefont {L.}~\bibnamefont {Ella}},
  \bibinfo {author} {\bibfnamefont {A.}~\bibnamefont {Rozen}}, \bibinfo
  {author} {\bibfnamefont {J.}~\bibnamefont {Birkbeck}}, \bibinfo {author}
  {\bibfnamefont {D.~J.}\ \bibnamefont {Perello}}, \bibinfo {author}
  {\bibfnamefont {D.}~\bibnamefont {Dutta}}, \bibinfo {author} {\bibfnamefont
  {M.}~\bibnamefont {Ben-Shalom}}, \bibinfo {author} {\bibfnamefont
  {T.}~\bibnamefont {Taniguchi}}, \bibinfo {author} {\bibfnamefont
  {K.}~\bibnamefont {Watanabe}}, \bibinfo {author} {\bibfnamefont
  {T.}~\bibnamefont {Holder}}, \bibinfo {author} {\bibfnamefont
  {R.}~\bibnamefont {Queiroz}}, \bibinfo {author} {\bibfnamefont
  {A.}~\bibnamefont {Principi}}, \bibinfo {author} {\bibfnamefont
  {A.}~\bibnamefont {Stern}}, \bibinfo {author} {\bibfnamefont
  {T.}~\bibnamefont {Scaffidi}}, \bibinfo {author} {\bibfnamefont {A.~K.}\
  \bibnamefont {Geim}},\ and\ \bibinfo {author} {\bibfnamefont
  {S.}~\bibnamefont {Ilani}},\ }\bibfield  {title} {\bibinfo {title}
  {{Visualizing Poiseuille flow of hydrodynamic electrons}},\ }\href
  {https://doi.org/10.1038/s41586-019-1788-9} {\bibfield  {journal} {\bibinfo
  {journal} {Nature (London)}\ }\textbf {\bibinfo {volume} {576}},\ \bibinfo
  {pages} {75} (\bibinfo {year} {2019})}\BibitemShut {NoStop}%
\bibitem [{\citenamefont {Lucas}\ and\ \citenamefont
  {Fong}(2017)}]{Lucas-Fong:rev-2017}%
  \BibitemOpen
  \bibfield  {author} {\bibinfo {author} {\bibfnamefont {A.}~\bibnamefont
  {Lucas}}\ and\ \bibinfo {author} {\bibfnamefont {K.~C.}\ \bibnamefont
  {Fong}},\ }\bibfield  {title} {\bibinfo {title} {{Hydrodynamics of electrons
  in graphene}},\ }\href {https://doi.org/10.1088/1361-648X/aaa274} {\bibfield
  {journal} {\bibinfo  {journal} {J. Phys. Condens. Matter}\ }\textbf {\bibinfo
  {volume} {94}},\ \bibinfo {pages} {2280} (\bibinfo {year}
  {2017})}\BibitemShut {NoStop}%
\bibitem [{\citenamefont {Narozhny}(2019)}]{Narozhny:rev-2019}%
  \BibitemOpen
  \bibfield  {author} {\bibinfo {author} {\bibfnamefont {B.~N.}\ \bibnamefont
  {Narozhny}},\ }\bibfield  {title} {\bibinfo {title} {{Electronic
  hydrodynamics in graphene}},\ }\href
  {https://doi.org/10.1016/j.aop.2019.167979} {\bibfield  {journal} {\bibinfo
  {journal} {Ann. Phys. (Amsterdam))}\ }\textbf {\bibinfo {volume} {411}},\
  \bibinfo {pages} {167979} (\bibinfo {year} {2019})}\BibitemShut {NoStop}%
\bibitem [{\citenamefont {Gooth}\ \emph {et~al.}(2018)\citenamefont {Gooth},
  \citenamefont {Menges}, \citenamefont {Kumar}, \citenamefont
  {S{\"{u}}$\beta$}, \citenamefont {Shekhar}, \citenamefont {Sun},
  \citenamefont {Drechsler}, \citenamefont {Zierold}, \citenamefont {Felser},\
  and\ \citenamefont {Gotsmann}}]{Gooth-Felser:2018}%
  \BibitemOpen
  \bibfield  {author} {\bibinfo {author} {\bibfnamefont {J.}~\bibnamefont
  {Gooth}}, \bibinfo {author} {\bibfnamefont {F.}~\bibnamefont {Menges}},
  \bibinfo {author} {\bibfnamefont {N.}~\bibnamefont {Kumar}}, \bibinfo
  {author} {\bibfnamefont {V.}~\bibnamefont {S{\"{u}}$\beta$}}, \bibinfo
  {author} {\bibfnamefont {C.}~\bibnamefont {Shekhar}}, \bibinfo {author}
  {\bibfnamefont {Y.}~\bibnamefont {Sun}}, \bibinfo {author} {\bibfnamefont
  {U.}~\bibnamefont {Drechsler}}, \bibinfo {author} {\bibfnamefont
  {R.}~\bibnamefont {Zierold}}, \bibinfo {author} {\bibfnamefont
  {C.}~\bibnamefont {Felser}},\ and\ \bibinfo {author} {\bibfnamefont
  {B.}~\bibnamefont {Gotsmann}},\ }\bibfield  {title} {\bibinfo {title}
  {{Thermal and electrical signatures of a hydrodynamic electron fluid in
  tungsten diphosphide}},\ }\href {https://doi.org/10.1038/s41467-018-06688-y}
  {\bibfield  {journal} {\bibinfo  {journal} {Nat. Commun.}\ }\textbf {\bibinfo
  {volume} {9}},\ \bibinfo {pages} {4093} (\bibinfo {year} {2018})}\BibitemShut
  {NoStop}%
\bibitem [{\citenamefont {Tomadin}\ and\ \citenamefont
  {Polini}(2013)}]{Tomadin-Polini:2013}%
  \BibitemOpen
  \bibfield  {author} {\bibinfo {author} {\bibfnamefont {A.}~\bibnamefont
  {Tomadin}}\ and\ \bibinfo {author} {\bibfnamefont {M.}~\bibnamefont
  {Polini}},\ }\bibfield  {title} {\bibinfo {title} {{Theory of the plasma-wave
  photoresponse of a gated graphene sheet}},\ }\href
  {https://doi.org/10.1103/PhysRevB.88.205426} {\bibfield  {journal} {\bibinfo
  {journal} {Phys. Rev. B}\ }\textbf {\bibinfo {volume} {88}},\ \bibinfo
  {pages} {205426} (\bibinfo {year} {2013})}\BibitemShut {NoStop}%
\bibitem [{\citenamefont {Svintsov}\ \emph {et~al.}(2013)\citenamefont
  {Svintsov}, \citenamefont {Vyurkov}, \citenamefont {Ryzhii},\ and\
  \citenamefont {Otsuji}}]{Svintsov-Otsuji:2013}%
  \BibitemOpen
  \bibfield  {author} {\bibinfo {author} {\bibfnamefont {D.}~\bibnamefont
  {Svintsov}}, \bibinfo {author} {\bibfnamefont {V.}~\bibnamefont {Vyurkov}},
  \bibinfo {author} {\bibfnamefont {V.}~\bibnamefont {Ryzhii}},\ and\ \bibinfo
  {author} {\bibfnamefont {T.}~\bibnamefont {Otsuji}},\ }\bibfield  {title}
  {\bibinfo {title} {{Hydrodynamic electron transport and nonlinear waves in
  graphene}},\ }\href {https://doi.org/10.1103/PhysRevB.88.245444} {\bibfield
  {journal} {\bibinfo  {journal} {Phys. Rev. B}\ }\textbf {\bibinfo {volume}
  {88}},\ \bibinfo {pages} {245444} (\bibinfo {year} {2013})}\BibitemShut
  {NoStop}%
\bibitem [{\citenamefont {Koseki}\ \emph {et~al.}(2016)\citenamefont {Koseki},
  \citenamefont {Ryzhii}, \citenamefont {Otsuji}, \citenamefont {Popov},\ and\
  \citenamefont {Satou}}]{Koseki-Satou:2016}%
  \BibitemOpen
  \bibfield  {author} {\bibinfo {author} {\bibfnamefont {Y.}~\bibnamefont
  {Koseki}}, \bibinfo {author} {\bibfnamefont {V.}~\bibnamefont {Ryzhii}},
  \bibinfo {author} {\bibfnamefont {T.}~\bibnamefont {Otsuji}}, \bibinfo
  {author} {\bibfnamefont {V.~V.}\ \bibnamefont {Popov}},\ and\ \bibinfo
  {author} {\bibfnamefont {A.}~\bibnamefont {Satou}},\ }\bibfield  {title}
  {\bibinfo {title} {{Giant plasmon instability in a dual-grating-gate graphene
  field-effect transistor}},\ }\href
  {https://doi.org/10.1103/PhysRevB.93.245408} {\bibfield  {journal} {\bibinfo
  {journal} {Phys. Rev. B}\ }\textbf {\bibinfo {volume} {93}},\ \bibinfo
  {pages} {245408} (\bibinfo {year} {2016})}\BibitemShut {NoStop}%
\bibitem [{\citenamefont {Mendl}\ \emph {et~al.}(2021)\citenamefont {Mendl},
  \citenamefont {Polini},\ and\ \citenamefont {Lucas}}]{Mendl-Polini:2019}%
  \BibitemOpen
  \bibfield  {author} {\bibinfo {author} {\bibfnamefont {C.~B.}\ \bibnamefont
  {Mendl}}, \bibinfo {author} {\bibfnamefont {M.}~\bibnamefont {Polini}},\ and\
  \bibinfo {author} {\bibfnamefont {A.}~\bibnamefont {Lucas}},\ }\bibfield
  {title} {\bibinfo {title} {Coherent terahertz radiation from a nonlinear
  oscillator of viscous electrons},\ }\href {https://doi.org/10.1063/5.0030869}
  {\bibfield  {journal} {\bibinfo  {journal} {Appl. Phys. Lett.}\ }\textbf
  {\bibinfo {volume} {118}},\ \bibinfo {pages} {013105} (\bibinfo {year}
  {2021})}\BibitemShut {NoStop}%
\bibitem [{\citenamefont {Crabb}\ \emph {et~al.}(2021)\citenamefont {Crabb},
  \citenamefont {Cantos-Roman}, \citenamefont {Jornet},\ and\ \citenamefont
  {Aizin}}]{Crabb-Aizin:2021}%
  \BibitemOpen
  \bibfield  {author} {\bibinfo {author} {\bibfnamefont {J.}~\bibnamefont
  {Crabb}}, \bibinfo {author} {\bibfnamefont {X.}~\bibnamefont {Cantos-Roman}},
  \bibinfo {author} {\bibfnamefont {J.~M.}\ \bibnamefont {Jornet}},\ and\
  \bibinfo {author} {\bibfnamefont {G.~R.}\ \bibnamefont {Aizin}},\ }\href@noop
  {} {\bibinfo {title} {{Hydrodynamic theory of the Dyakonov-Shur instability
  in graphene transistors}}} (\bibinfo {year} {2021}),\ \Eprint
  {https://arxiv.org/abs/2106.01296} {arXiv:2106.01296} \BibitemShut {NoStop}%
\bibitem [{\citenamefont {Tauk}\ \emph {et~al.}(2006)\citenamefont {Tauk},
  \citenamefont {Teppe}, \citenamefont {Boubanga}, \citenamefont {Coquillat},
  \citenamefont {Knap}, \citenamefont {Meziani}, \citenamefont {Gallon},
  \citenamefont {Boeuf}, \citenamefont {Skotnicki}, \citenamefont
  {Fenouillet-Beranger}, \citenamefont {Maude}, \citenamefont {Rumyantsev},\
  and\ \citenamefont {Shur}}]{Tauk-Shur:2006}%
  \BibitemOpen
  \bibfield  {author} {\bibinfo {author} {\bibfnamefont {R.}~\bibnamefont
  {Tauk}}, \bibinfo {author} {\bibfnamefont {F.}~\bibnamefont {Teppe}},
  \bibinfo {author} {\bibfnamefont {S.}~\bibnamefont {Boubanga}}, \bibinfo
  {author} {\bibfnamefont {D.}~\bibnamefont {Coquillat}}, \bibinfo {author}
  {\bibfnamefont {W.}~\bibnamefont {Knap}}, \bibinfo {author} {\bibfnamefont
  {Y.~M.}\ \bibnamefont {Meziani}}, \bibinfo {author} {\bibfnamefont
  {C.}~\bibnamefont {Gallon}}, \bibinfo {author} {\bibfnamefont
  {F.}~\bibnamefont {Boeuf}}, \bibinfo {author} {\bibfnamefont
  {T.}~\bibnamefont {Skotnicki}}, \bibinfo {author} {\bibfnamefont
  {C.}~\bibnamefont {Fenouillet-Beranger}}, \bibinfo {author} {\bibfnamefont
  {D.~K.}\ \bibnamefont {Maude}}, \bibinfo {author} {\bibfnamefont
  {S.}~\bibnamefont {Rumyantsev}},\ and\ \bibinfo {author} {\bibfnamefont
  {M.~S.}\ \bibnamefont {Shur}},\ }\bibfield  {title} {\bibinfo {title}
  {{Plasma wave detection of terahertz radiation by silicon field effects
  transistors: Responsivity and noise equivalent power}},\ }\href
  {https://doi.org/10.1063/1.2410215} {\bibfield  {journal} {\bibinfo
  {journal} {Appl. Phys. Lett.}\ }\textbf {\bibinfo {volume} {89}},\ \bibinfo
  {pages} {253511} (\bibinfo {year} {2006})}\BibitemShut {NoStop}%
\bibitem [{\citenamefont {Vitiello}\ \emph {et~al.}(2012)\citenamefont
  {Vitiello}, \citenamefont {Coquillat}, \citenamefont {Viti}, \citenamefont
  {Ercolani}, \citenamefont {Teppe}, \citenamefont {Pitanti}, \citenamefont
  {Beltram}, \citenamefont {Sorba}, \citenamefont {Knap},\ and\ \citenamefont
  {Tredicucci}}]{Vitiello-Tredicucci:2012}%
  \BibitemOpen
  \bibfield  {author} {\bibinfo {author} {\bibfnamefont {M.~S.}\ \bibnamefont
  {Vitiello}}, \bibinfo {author} {\bibfnamefont {D.}~\bibnamefont {Coquillat}},
  \bibinfo {author} {\bibfnamefont {L.}~\bibnamefont {Viti}}, \bibinfo {author}
  {\bibfnamefont {D.}~\bibnamefont {Ercolani}}, \bibinfo {author}
  {\bibfnamefont {F.}~\bibnamefont {Teppe}}, \bibinfo {author} {\bibfnamefont
  {A.}~\bibnamefont {Pitanti}}, \bibinfo {author} {\bibfnamefont
  {F.}~\bibnamefont {Beltram}}, \bibinfo {author} {\bibfnamefont
  {L.}~\bibnamefont {Sorba}}, \bibinfo {author} {\bibfnamefont
  {W.}~\bibnamefont {Knap}},\ and\ \bibinfo {author} {\bibfnamefont
  {A.}~\bibnamefont {Tredicucci}},\ }\bibfield  {title} {\bibinfo {title}
  {{Room-temperature terahertz detectors based on semiconductor nanowire
  field-effect transistors}},\ }\href {https://doi.org/10.1021/nl2030486}
  {\bibfield  {journal} {\bibinfo  {journal} {Nano Lett.}\ }\textbf {\bibinfo
  {volume} {12}},\ \bibinfo {pages} {96} (\bibinfo {year} {2012})}\BibitemShut
  {NoStop}%
\bibitem [{\citenamefont {Vicarelli}\ \emph {et~al.}(2012)\citenamefont
  {Vicarelli}, \citenamefont {Vitiello}, \citenamefont {Coquillat},
  \citenamefont {Lombardo}, \citenamefont {Ferrari}, \citenamefont {Knap},
  \citenamefont {Polini}, \citenamefont {Pellegrini},\ and\ \citenamefont
  {Tredicucci}}]{Vicarelli-Polini-Tredicucci:2012}%
  \BibitemOpen
  \bibfield  {author} {\bibinfo {author} {\bibfnamefont {L.}~\bibnamefont
  {Vicarelli}}, \bibinfo {author} {\bibfnamefont {M.~S.}\ \bibnamefont
  {Vitiello}}, \bibinfo {author} {\bibfnamefont {D.}~\bibnamefont {Coquillat}},
  \bibinfo {author} {\bibfnamefont {A.}~\bibnamefont {Lombardo}}, \bibinfo
  {author} {\bibfnamefont {A.~C.}\ \bibnamefont {Ferrari}}, \bibinfo {author}
  {\bibfnamefont {W.}~\bibnamefont {Knap}}, \bibinfo {author} {\bibfnamefont
  {M.}~\bibnamefont {Polini}}, \bibinfo {author} {\bibfnamefont
  {V.}~\bibnamefont {Pellegrini}},\ and\ \bibinfo {author} {\bibfnamefont
  {A.}~\bibnamefont {Tredicucci}},\ }\bibfield  {title} {\bibinfo {title}
  {{Graphene field effect transistors as room-temperature terahertz
  detectors}},\ }\href {https://doi.org/10.1038/nmat3417} {\bibfield  {journal}
  {\bibinfo  {journal} {Nat. Mater.}\ }\textbf {\bibinfo {volume} {11}},\
  \bibinfo {pages} {865} (\bibinfo {year} {2012})}\BibitemShut {NoStop}%
\bibitem [{\citenamefont {Giliberti}\ \emph {et~al.}(2015)\citenamefont
  {Giliberti}, \citenamefont {{Di Gaspare}}, \citenamefont {Giovine},
  \citenamefont {Ortolani}, \citenamefont {Sorba}, \citenamefont {Biasiol},
  \citenamefont {Popov}, \citenamefont {Fateev},\ and\ \citenamefont
  {Evangelisti}}]{Giliberti-Evangelisti:2015}%
  \BibitemOpen
  \bibfield  {author} {\bibinfo {author} {\bibfnamefont {V.}~\bibnamefont
  {Giliberti}}, \bibinfo {author} {\bibfnamefont {A.}~\bibnamefont {{Di
  Gaspare}}}, \bibinfo {author} {\bibfnamefont {E.}~\bibnamefont {Giovine}},
  \bibinfo {author} {\bibfnamefont {M.}~\bibnamefont {Ortolani}}, \bibinfo
  {author} {\bibfnamefont {L.}~\bibnamefont {Sorba}}, \bibinfo {author}
  {\bibfnamefont {G.}~\bibnamefont {Biasiol}}, \bibinfo {author} {\bibfnamefont
  {V.~V.}\ \bibnamefont {Popov}}, \bibinfo {author} {\bibfnamefont {D.~V.}\
  \bibnamefont {Fateev}},\ and\ \bibinfo {author} {\bibfnamefont
  {F.}~\bibnamefont {Evangelisti}},\ }\bibfield  {title} {\bibinfo {title}
  {{Downconversion of terahertz radiation due to intrinsic hydrodynamic
  nonlinearity of a two-dimensional electron plasma}},\ }\href
  {https://doi.org/10.1103/PhysRevB.91.165313} {\bibfield  {journal} {\bibinfo
  {journal} {Phys. Rev. B}\ }\textbf {\bibinfo {volume} {91}},\ \bibinfo
  {pages} {165313} (\bibinfo {year} {2015})}\BibitemShut {NoStop}%
\bibitem [{\citenamefont {Bandurin}\ \emph
  {et~al.}(2018{\natexlab{b}})\citenamefont {Bandurin}, \citenamefont
  {Gayduchenko}, \citenamefont {Cao}, \citenamefont {Moskotin}, \citenamefont
  {Principi}, \citenamefont {Grigorieva}, \citenamefont {Goltsman},
  \citenamefont {Fedorov},\ and\ \citenamefont
  {Svintsov}}]{Bandurin-Svintsov:2018}%
  \BibitemOpen
  \bibfield  {author} {\bibinfo {author} {\bibfnamefont {D.~A.}\ \bibnamefont
  {Bandurin}}, \bibinfo {author} {\bibfnamefont {I.}~\bibnamefont
  {Gayduchenko}}, \bibinfo {author} {\bibfnamefont {Y.}~\bibnamefont {Cao}},
  \bibinfo {author} {\bibfnamefont {M.}~\bibnamefont {Moskotin}}, \bibinfo
  {author} {\bibfnamefont {A.}~\bibnamefont {Principi}}, \bibinfo {author}
  {\bibfnamefont {I.~V.}\ \bibnamefont {Grigorieva}}, \bibinfo {author}
  {\bibfnamefont {G.}~\bibnamefont {Goltsman}}, \bibinfo {author}
  {\bibfnamefont {G.}~\bibnamefont {Fedorov}},\ and\ \bibinfo {author}
  {\bibfnamefont {D.}~\bibnamefont {Svintsov}},\ }\bibfield  {title} {\bibinfo
  {title} {{Dual origin of room temperature sub-terahertz photoresponse in
  graphene field effect transistors}},\ }\href
  {https://doi.org/10.1063/1.5018151} {\bibfield  {journal} {\bibinfo
  {journal} {Appl. Phys. Lett.}\ }\textbf {\bibinfo {volume} {112}},\ \bibinfo
  {pages} {141101} (\bibinfo {year} {2018}{\natexlab{b}})}\BibitemShut
  {NoStop}%
\bibitem [{\citenamefont {Landau}\ and\ \citenamefont
  {Lifshitz}(2013)}]{Landau:t6-2013}%
  \BibitemOpen
  \bibfield  {author} {\bibinfo {author} {\bibfnamefont {L.~D.}\ \bibnamefont
  {Landau}}\ and\ \bibinfo {author} {\bibfnamefont {E.~M.}\ \bibnamefont
  {Lifshitz}},\ }\href
  {https://www.elsevier.com/books/fluid-mechanics/landau/978-0-08-057073-0}
  {\emph {\bibinfo {title} {{Fluid Mechanics}}}}\ (\bibinfo  {publisher}
  {Butterworth-Heinemann},\ \bibinfo {address} {Oxford},\ \bibinfo {year}
  {2013})\BibitemShut {NoStop}%
\bibitem [{\citenamefont {Ogilvie}(2016)}]{Ogilvie:2016}%
  \BibitemOpen
  \bibfield  {author} {\bibinfo {author} {\bibfnamefont {G.~I.}\ \bibnamefont
  {Ogilvie}},\ }\bibfield  {title} {\bibinfo {title} {{Astrophysical fluid
  dynamics}},\ }\href {https://doi.org/10.1017/S0022377816000489} {\bibfield
  {journal} {\bibinfo  {journal} {J. Plasma Phys.}\ }\textbf {\bibinfo {volume}
  {82}},\ \bibinfo {pages} {205820301} (\bibinfo {year} {2016})}\BibitemShut
  {NoStop}%
\bibitem [{SM()}]{SM}%
  \BibitemOpen
  \href@noop {} {\bibinfo  {journal} {See Supplemental Material for details of
  the derivations of thermodynamic variables, boundary conditions, solutions
  for the hydrodynamic deviations, and frequencies of plasmons and entropy
  modes in graphene. The Supplemental Material contains
  Refs.~\cite{Gantmakher-Levinson:book,Gorbar:2017vph,Abedinpour-MacDonald:2011}}\
  }\BibitemShut {NoStop}%
\bibitem [{Note1()}]{Note1}%
  \BibitemOpen
\bibfield  {journal} {  }\bibinfo {note} {The bulk viscosity is vanishingly
  small in systems with relativisticlike quasiparticles, see, e.g., Ref.~\cite
  {Principi-Polini:2015} for an explicit calculation in graphene.}\BibitemShut
  {Stop}%
\bibitem [{\citenamefont {Shur}(1987)}]{Shur:book}%
  \BibitemOpen
  \bibfield  {author} {\bibinfo {author} {\bibfnamefont {M.~S.}\ \bibnamefont
  {Shur}},\ }\href {https://doi.org/10.1007/978-1-4899-1989-2} {\emph {\bibinfo
  {title} {{GaAs Devices and Circuits}}}}\ (\bibinfo  {publisher} {Springer
  US},\ \bibinfo {address} {New York},\ \bibinfo {year} {1987})\BibitemShut
  {NoStop}%
\bibitem [{\citenamefont {Rossby}(1939)}]{Rossby:1938}%
  \BibitemOpen
  \bibfield  {author} {\bibinfo {author} {\bibfnamefont {C.-G.}\ \bibnamefont
  {Rossby}},\ }\bibfield  {title} {\bibinfo {title} {{Relation between
  variations in the intensity of the zonal circulation of the atmosphere and
  the displacements of the semi-permanent centers of action}},\ }\href@noop {}
  {\bibfield  {journal} {\bibinfo  {journal} {J. Mar. Res.}\ }\textbf {\bibinfo
  {volume} {2}},\ \bibinfo {pages} {38} (\bibinfo {year} {1939})}\BibitemShut
  {NoStop}%
\bibitem [{\citenamefont {Gill}(1982)}]{Gill:book}%
  \BibitemOpen
  \bibfield  {author} {\bibinfo {author} {\bibfnamefont {A.}~\bibnamefont
  {Gill}},\ }\href@noop {} {\emph {\bibinfo {title} {{Atmosphere-Ocean
  Dynamics}}}}\ (\bibinfo  {publisher} {Academic Press},\ \bibinfo {address}
  {New York},\ \bibinfo {year} {1982})\BibitemShut {NoStop}%
\bibitem [{Note2()}]{Note2}%
  \BibitemOpen
  \bibinfo {note} {The growth rate for plasmons (\ref {DS-inf-omega-pm-2D})
  differs from that in Ref.~\cite {Tomadin-Polini:2013}. The discrepancy can be
  traced to a different form of the nonlinear terms in the Navier-Stokes
  equation and the dependence of pressure $P$ on velocity.}\BibitemShut {Stop}%
\bibitem [{Note3()}]{Note3}%
  \BibitemOpen
  \bibinfo {note} {The hydrodynamic description is invalid for $L \omega
  _p/v_s\lesssim 1$ because the system size becomes comparable to the
  inter-carrier distance in this case.}\BibitemShut {Stop}%
\bibitem [{\citenamefont {Kumar}\ \emph {et~al.}(2017)\citenamefont {Kumar},
  \citenamefont {Sun}, \citenamefont {Xu}, \citenamefont {Manna}, \citenamefont
  {Yao}, \citenamefont {S{\"{u}}ss}, \citenamefont {Leermakers}, \citenamefont
  {Young}, \citenamefont {F{\"{o}}rster}, \citenamefont {Schmidt},
  \citenamefont {Borrmann}, \citenamefont {Yan}, \citenamefont {Zeitler},
  \citenamefont {Shi}, \citenamefont {Felser},\ and\ \citenamefont
  {Shekhar}}]{Kumar-Felser:2017}%
  \BibitemOpen
  \bibfield  {author} {\bibinfo {author} {\bibfnamefont {N.}~\bibnamefont
  {Kumar}}, \bibinfo {author} {\bibfnamefont {Y.}~\bibnamefont {Sun}}, \bibinfo
  {author} {\bibfnamefont {N.}~\bibnamefont {Xu}}, \bibinfo {author}
  {\bibfnamefont {K.}~\bibnamefont {Manna}}, \bibinfo {author} {\bibfnamefont
  {M.}~\bibnamefont {Yao}}, \bibinfo {author} {\bibfnamefont {V.}~\bibnamefont
  {S{\"{u}}ss}}, \bibinfo {author} {\bibfnamefont {I.}~\bibnamefont
  {Leermakers}}, \bibinfo {author} {\bibfnamefont {O.}~\bibnamefont {Young}},
  \bibinfo {author} {\bibfnamefont {T.}~\bibnamefont {F{\"{o}}rster}}, \bibinfo
  {author} {\bibfnamefont {M.}~\bibnamefont {Schmidt}}, \bibinfo {author}
  {\bibfnamefont {H.}~\bibnamefont {Borrmann}}, \bibinfo {author}
  {\bibfnamefont {B.}~\bibnamefont {Yan}}, \bibinfo {author} {\bibfnamefont
  {U.}~\bibnamefont {Zeitler}}, \bibinfo {author} {\bibfnamefont
  {M.}~\bibnamefont {Shi}}, \bibinfo {author} {\bibfnamefont {C.}~\bibnamefont
  {Felser}},\ and\ \bibinfo {author} {\bibfnamefont {C.}~\bibnamefont
  {Shekhar}},\ }\bibfield  {title} {\bibinfo {title} {{Extremely high
  magnetoresistance and conductivity in the type-II Weyl semimetals WP$_2$ and
  MoP$_2$}},\ }\href {https://doi.org/10.1038/s41467-017-01758-z} {\bibfield
  {journal} {\bibinfo  {journal} {Nat. Commun.}\ }\textbf {\bibinfo {volume}
  {8}},\ \bibinfo {pages} {1642} (\bibinfo {year} {2017})}\BibitemShut
  {NoStop}%
\bibitem [{\citenamefont {Gantmakher}\ and\ \citenamefont
  {Levinson}(1987)}]{Gantmakher-Levinson:book}%
  \BibitemOpen
  \bibfield  {author} {\bibinfo {author} {\bibfnamefont {V.~F.}\ \bibnamefont
  {Gantmakher}}\ and\ \bibinfo {author} {\bibfnamefont {Y.~B.}\ \bibnamefont
  {Levinson}},\ }\href
  {https://www.elsevier.com/books/carrier-scattering-in-metals-and-semiconductors/gantmakher/978-0-444-87025-4}
  {\emph {\bibinfo {title} {{Carrier scattering in metals and
  semiconductors}}}},\ Modern problems in condensed matter sciences\ (\bibinfo
  {publisher} {North-Holland},\ \bibinfo {address} {Amsterdam},\ \bibinfo
  {year} {1987})\BibitemShut {NoStop}%
\bibitem [{\citenamefont {Gorbar}\ \emph {et~al.}(2018)\citenamefont {Gorbar},
  \citenamefont {Miransky}, \citenamefont {Shovkovy},\ and\ \citenamefont
  {Sukhachov}}]{Gorbar:2017vph}%
  \BibitemOpen
  \bibfield  {author} {\bibinfo {author} {\bibfnamefont {E.~V.}\ \bibnamefont
  {Gorbar}}, \bibinfo {author} {\bibfnamefont {V.~A.}\ \bibnamefont
  {Miransky}}, \bibinfo {author} {\bibfnamefont {I.~A.}\ \bibnamefont
  {Shovkovy}},\ and\ \bibinfo {author} {\bibfnamefont {P.~O.}\ \bibnamefont
  {Sukhachov}},\ }\bibfield  {title} {\bibinfo {title} {{Consistent
  hydrodynamic theory of chiral electrons in Weyl semimetals}},\ }\href
  {https://doi.org/10.1103/PhysRevB.97.121105} {\bibfield  {journal} {\bibinfo
  {journal} {Phys. Rev. B}\ }\textbf {\bibinfo {volume} {97}},\ \bibinfo
  {pages} {121105(R)} (\bibinfo {year} {2018})}\BibitemShut {NoStop}%
\bibitem [{\citenamefont {Abedinpour}\ \emph {et~al.}(2011)\citenamefont
  {Abedinpour}, \citenamefont {Vignale}, \citenamefont {Principi},
  \citenamefont {Polini}, \citenamefont {Tse},\ and\ \citenamefont
  {MacDonald}}]{Abedinpour-MacDonald:2011}%
  \BibitemOpen
  \bibfield  {author} {\bibinfo {author} {\bibfnamefont {S.~H.}\ \bibnamefont
  {Abedinpour}}, \bibinfo {author} {\bibfnamefont {G.}~\bibnamefont {Vignale}},
  \bibinfo {author} {\bibfnamefont {A.}~\bibnamefont {Principi}}, \bibinfo
  {author} {\bibfnamefont {M.}~\bibnamefont {Polini}}, \bibinfo {author}
  {\bibfnamefont {W.-K.}\ \bibnamefont {Tse}},\ and\ \bibinfo {author}
  {\bibfnamefont {A.~H.}\ \bibnamefont {MacDonald}},\ }\bibfield  {title}
  {\bibinfo {title} {{Drude weight, plasmon dispersion, and ac conductivity in
  doped graphene sheets}},\ }\href {https://doi.org/10.1103/PhysRevB.84.045429}
  {\bibfield  {journal} {\bibinfo  {journal} {Phys. Rev. B}\ }\textbf {\bibinfo
  {volume} {84}},\ \bibinfo {pages} {045429} (\bibinfo {year}
  {2011})}\BibitemShut {NoStop}%
\bibitem [{\citenamefont {Principi}\ \emph {et~al.}(2016)\citenamefont
  {Principi}, \citenamefont {Vignale}, \citenamefont {Carrega},\ and\
  \citenamefont {Polini}}]{Principi-Polini:2015}%
  \BibitemOpen
  \bibfield  {author} {\bibinfo {author} {\bibfnamefont {A.}~\bibnamefont
  {Principi}}, \bibinfo {author} {\bibfnamefont {G.}~\bibnamefont {Vignale}},
  \bibinfo {author} {\bibfnamefont {M.}~\bibnamefont {Carrega}},\ and\ \bibinfo
  {author} {\bibfnamefont {M.}~\bibnamefont {Polini}},\ }\bibfield  {title}
  {\bibinfo {title} {{Bulk and shear viscosities of the two-dimensional
  electron liquid in a doped graphene sheet}},\ }\href
  {https://doi.org/10.1103/PhysRevB.93.125410} {\bibfield  {journal} {\bibinfo
  {journal} {Phys. Rev. B}\ }\textbf {\bibinfo {volume} {93}},\ \bibinfo
  {pages} {125410} (\bibinfo {year} {2016})}\BibitemShut {NoStop}%
\end{thebibliography}%


\begin{thebibliography}{6}%
\makeatletter
\providecommand \@ifxundefined [1]{%
 \@ifx{#1\undefined}
}%
\providecommand \@ifnum [1]{%
 \ifnum #1\expandafter \@firstoftwo
 \else \expandafter \@secondoftwo
 \fi
}%
\providecommand \@ifx [1]{%
 \ifx #1\expandafter \@firstoftwo
 \else \expandafter \@secondoftwo
 \fi
}%
\providecommand \natexlab [1]{#1}%
\providecommand \enquote  [1]{``#1''}%
\providecommand \bibnamefont  [1]{#1}%
\providecommand \bibfnamefont [1]{#1}%
\providecommand \citenamefont [1]{#1}%
\providecommand \href@noop [0]{\@secondoftwo}%
\providecommand \href [0]{\begingroup \@sanitize@url \@href}%
\providecommand \@href[1]{\@@startlink{#1}\@@href}%
\providecommand \@@href[1]{\endgroup#1\@@endlink}%
\providecommand \@sanitize@url [0]{\catcode `\\12\catcode `\$12\catcode
  `\&12\catcode `\#12\catcode `\^12\catcode `\_12\catcode `\%12\relax}%
\providecommand \@@startlink[1]{}%
\providecommand \@@endlink[0]{}%
\providecommand \url  [0]{\begingroup\@sanitize@url \@url }%
\providecommand \@url [1]{\endgroup\@href {#1}{\urlprefix }}%
\providecommand \urlprefix  [0]{URL }%
\providecommand \Eprint [0]{\href }%
\providecommand \doibase [0]{https://doi.org/}%
\providecommand \selectlanguage [0]{\@gobble}%
\providecommand \bibinfo  [0]{\@secondoftwo}%
\providecommand \bibfield  [0]{\@secondoftwo}%
\providecommand \translation [1]{[#1]}%
\providecommand \BibitemOpen [0]{}%
\providecommand \bibitemStop [0]{}%
\providecommand \bibitemNoStop [0]{.\EOS\space}%
\providecommand \EOS [0]{\spacefactor3000\relax}%
\providecommand \BibitemShut  [1]{\csname bibitem#1\endcsname}%
\let\auto@bib@innerbib\@empty
\bibitem [{\citenamefont {Gantmakher}\ and\ \citenamefont
  {Levinson}(1987)}]{Gantmakher-Levinson:book}%
  \BibitemOpen
  \bibfield  {author} {\bibinfo {author} {\bibfnamefont {V.~F.}\ \bibnamefont
  {Gantmakher}}\ and\ \bibinfo {author} {\bibfnamefont {Y.~B.}\ \bibnamefont
  {Levinson}},\ }\href
  {https://www.elsevier.com/books/carrier-scattering-in-metals-and-semiconductors/gantmakher/978-0-444-87025-4}
  {\emph {\bibinfo {title} {{Carrier scattering in metals and
  semiconductors}}}},\ Modern problems in condensed matter sciences\ (\bibinfo
  {publisher} {North-Holland},\ \bibinfo {address} {Amsterdam},\ \bibinfo
  {year} {1987})\BibitemShut {NoStop}%
\bibitem [{\citenamefont {Lucas}\ and\ \citenamefont
  {Fong}(2017)}]{Lucas-Fong:rev-2017}%
  \BibitemOpen
  \bibfield  {author} {\bibinfo {author} {\bibfnamefont {A.}~\bibnamefont
  {Lucas}}\ and\ \bibinfo {author} {\bibfnamefont {K.~C.}\ \bibnamefont
  {Fong}},\ }\bibfield  {title} {\bibinfo {title} {{Hydrodynamics of electrons
  in graphene}},\ }\href {https://doi.org/10.1088/1361-648X/aaa274} {\bibfield
  {journal} {\bibinfo  {journal} {J. Phys. Condens. Matter}\ }\textbf {\bibinfo
  {volume} {94}},\ \bibinfo {pages} {2280} (\bibinfo {year}
  {2017})}\BibitemShut {NoStop}%
\bibitem [{\citenamefont {Narozhny}(2019)}]{Narozhny:rev-2019}%
  \BibitemOpen
  \bibfield  {author} {\bibinfo {author} {\bibfnamefont {B.~N.}\ \bibnamefont
  {Narozhny}},\ }\bibfield  {title} {\bibinfo {title} {{Electronic
  hydrodynamics in graphene}},\ }\href
  {https://doi.org/10.1016/j.aop.2019.167979} {\bibfield  {journal} {\bibinfo
  {journal} {Ann. Phys. (Amsterdam))}\ }\textbf {\bibinfo {volume} {411}},\
  \bibinfo {pages} {167979} (\bibinfo {year} {2019})}\BibitemShut {NoStop}%
\bibitem [{\citenamefont {Gorbar}\ \emph {et~al.}(2018)\citenamefont {Gorbar},
  \citenamefont {Miransky}, \citenamefont {Shovkovy},\ and\ \citenamefont
  {Sukhachov}}]{Gorbar:2017vph}%
  \BibitemOpen
  \bibfield  {author} {\bibinfo {author} {\bibfnamefont {E.~V.}\ \bibnamefont
  {Gorbar}}, \bibinfo {author} {\bibfnamefont {V.~A.}\ \bibnamefont
  {Miransky}}, \bibinfo {author} {\bibfnamefont {I.~A.}\ \bibnamefont
  {Shovkovy}},\ and\ \bibinfo {author} {\bibfnamefont {P.~O.}\ \bibnamefont
  {Sukhachov}},\ }\bibfield  {title} {\bibinfo {title} {{Consistent
  hydrodynamic theory of chiral electrons in Weyl semimetals}},\ }\href
  {https://doi.org/10.1103/PhysRevB.97.121105} {\bibfield  {journal} {\bibinfo
  {journal} {Phys. Rev. B}\ }\textbf {\bibinfo {volume} {97}},\ \bibinfo
  {pages} {121105(R)} (\bibinfo {year} {2018})}\BibitemShut {NoStop}%
\bibitem [{\citenamefont {Shur}(1987)}]{Shur:book}%
  \BibitemOpen
  \bibfield  {author} {\bibinfo {author} {\bibfnamefont {M.~S.}\ \bibnamefont
  {Shur}},\ }\href {https://doi.org/10.1007/978-1-4899-1989-2} {\emph {\bibinfo
  {title} {{GaAs Devices and Circuits}}}}\ (\bibinfo  {publisher} {Springer
  US},\ \bibinfo {address} {New York},\ \bibinfo {year} {1987})\BibitemShut
  {NoStop}%
\bibitem [{\citenamefont {Dyakonov}\ and\ \citenamefont
  {Shur}(1993)}]{Dyakonov-Shur:1993}%
  \BibitemOpen
  \bibfield  {author} {\bibinfo {author} {\bibfnamefont {M.}~\bibnamefont
  {Dyakonov}}\ and\ \bibinfo {author} {\bibfnamefont {M.}~\bibnamefont
  {Shur}},\ }\bibfield  {title} {\bibinfo {title} {{Shallow water analogy for a
  ballistic field effect transistor: New mechanism of plasma wave generation by
  dc current}},\ }\href {https://doi.org/10.1103/PhysRevLett.71.2465}
  {\bibfield  {journal} {\bibinfo  {journal} {Phys. Rev. Lett.}\ }\textbf
  {\bibinfo {volume} {71}},\ \bibinfo {pages} {2465} (\bibinfo {year}
  {1993})}\BibitemShut {NoStop}%
\end{thebibliography}%

\end{document}


\title{Supplemental Material\\
Entropy Wave Instability in Dirac and Weyl Semimetals}

\author{P.~O.~Sukhachov}
\email{pavlo.sukhachov@yale.edu}
\affiliation{Department of Physics, Yale University, New Haven, Connecticut 06520, USA}

\author{E. V. Gorbar}
\affiliation{Department of Physics, Taras Shevchenko National University of Kyiv, Kyiv, 03022, Ukraine}
\affiliation{Bogolyubov Institute for Theoretical Physics, Kyiv, 03143, Ukraine}

\author{I.~A.~Shovkovy}
\affiliation{College of Integrative Sciences and Arts, Arizona State University, Mesa, Arizona 85212, USA}
\affiliation{Department of Physics, Arizona State University, Tempe, Arizona 85287, USA}

\maketitle
\tableofcontents

\section{S I. Thermodynamic variables}
\label{sec:app-def}

In this section, we sketch the derivation of the hydrodynamic equations used in the main text and present thermodynamic variables utilized in our analysis of instabilities. We use the following hydrodynamic ansatz for the distribution function~\cite{Gantmakher-Levinson:book,Lucas-Fong:rev-2017,Narozhny:rev-2019}:
\begin{equation}
\label{app-def-f-def}
f = \frac{1}{1+e^{\left[\epsilon-\left(\mathbf{u}\cdot\mathbf{p}\right) -\mu\right]/T}},
\end{equation}
where $\epsilon=v_F p$ for the relativisticlike spectrum, $v_F$ is the Fermi velocity, $p$ is the momentum, $\mu$ is the electric chemical potential, and $T$ is temperature. This distribution function satisfies identically the electron-electron collision integral.

In terms of the distribution function (\ref{app-def-f-def}), the electric charge density, the energy density, and the momentum flux tensor read as
\begin{eqnarray}
\label{app-def-n}
-en &=& -\sum_{\lambda}\sum_{\rm e,h}e\int\frac{d^dp}{(2\pi \hbar)^d} f,\\
\label{app-def-eps}
\epsilon &=& \sum_{\lambda}\sum_{\rm e,h}\int\frac{d^dp}{(2\pi \hbar)^d} v_F p f,\\
\label{app-def-Pij}
\Pi_{ij} &=&\sum_{\lambda}\sum_{\rm e,h}\int\frac{d^dp}{(2\pi \hbar)^d} v_F p_i \hat{\mathbf{p}}_j f,
\end{eqnarray}
where $\sum_{\lambda}$ denotes the summation over all Weyl nodes/spin degrees of freedom, $\sum_{\rm e,h}$ stands for the summation over electron and holes, and $d=2,3$ is the spatial dimension.

The hydrodynamic equations are derived by calculating the moments of the Boltzmann equation, i.e., by multiplying it by $\mathbf{p}$ and $\epsilon$ for the momentum conservation and energy continuity equations, respectively, and integrating over $\mathbf{p}$. The charge conservation relation is derived by simply integrating over momenta. The derivation is straightforward and can be found in Refs.~\cite{Lucas-Fong:rev-2017,Gorbar:2017vph,Narozhny:rev-2019}. Therefore, we do not present it here.

\subsection{S I.A. Comoving frame}
\label{sec:app-def-com}

Let us first calculate the thermodynamic variables in the comoving frame. For this, we set $u=0$ in Eq.~(\ref{app-def-f-def}) and use Eqs.~(\ref{app-def-n}) through (\ref{app-def-Pij}).

In 3D, the electron number density $n_{\rm eq}$ equals
\begin{equation}
\label{app-def-com-n}
n_{\rm eq} = -N_{W}\frac{T^{3}}{\pi^2 \hbar^3 v_F^{3}}  \left[\mbox{Li}_{3}\left(-e^{\mu/T}\right) -\mbox{Li}_{3}\left(-e^{-\mu/T}\right)\right] = N_{W}\frac{\mu \left(\mu^2 +\pi^2T^2\right)}{6\pi^2 v_F^3 \hbar^3},
\end{equation}
where $\mbox{Li}_{n}(x)$ is the polylogarithm function. The energy density $\epsilon_{\rm eq}$ reads
\begin{equation}
\label{app-def-com-eps}
\epsilon_{\rm eq} =-N_{W}\frac{3T^{4}}{\pi^2 \hbar^3 v_F^{\revisionC{3}}}  \left[\mbox{Li}_{4}\left(-e^{\mu/T}\right) +\mbox{Li}_{4}\left(-e^{-\mu/T}\right)\right]= N_{W}\frac{1}{8\pi^2 \hbar^3v_F^3} \left(\mu^4 +2\pi^2T^2\mu^2 +\frac{7\pi^4T^4}{15}\right).
\end{equation}
The prefactor $N_{W}$ in Eqs.~(\ref{app-def-com-n}) and (\ref{app-def-com-eps}) is the number of Weyl nodes. We have $N_{W}=2$ in 3D Dirac semimetals with a single Dirac point.

The electron number density $n_{\rm eq}$ and the energy density $\epsilon_{\rm eq}$ for a 2D relativisticlike spectrum are
\begin{equation}
\label{app-def-com-n-2D}
n_{\rm eq} = -N_g \frac{T^2}{2\pi v_F^2 \hbar^2} \left[\mbox{Li}_{2}\left(-e^{\mu/T}\right) -\mbox{Li}_{2}\left(-e^{-\mu/T}\right)\right]
\end{equation}
and
\begin{equation}
\label{app-def-com-eps-2D}
\epsilon_{\rm eq} = -N_g \frac{T^3}{\pi v_F^2 \hbar^2} \left[\mbox{Li}_{3}\left(-e^{\mu/T}\right) +\mbox{Li}_{3}\left(-e^{-\mu/T}\right)\right],
\end{equation}
respectively. Here, $N_g$ is the degeneracy factor. In graphene, $N_g=4$ accounts for the valley and spin degeneracy.

Pressure $P_{\rm eq}$ and the enthalpy density $w_{\rm eq}=\epsilon_{\rm eq}+P_{\rm eq}$ are given by the standard expressions for relativisticlike systems, $P_{\rm eq}=\epsilon_{\rm eq}/d$ and $w_{\rm eq}=(d+1)\epsilon_{\rm eq}/d$. Pressure $P_{\rm eq}$ is defined as a diagonal component of momentum flux tensor $\Pi_{{\rm eq}, ii}$. Notice also that the following relations are valid:
\begin{equation}
\label{app-def-com-eps-mu}
\partial_{\mu}\epsilon_{\rm eq} = n_{\rm eq} d, \quad \partial_{T}\epsilon_{\rm eq} = s_{\rm eq} d,
\end{equation}
where $s_{\rm eq}=(w_{\rm eq}-\mu n_{\rm eq})/T$ is the entropy density.

\subsection{S I.B. Laboratory frame}
\label{sec:app-def-lab}

In this section, we calculate the electric charge density, the energy density, and the momentum flux tensor in the laboratory frame where $u\neq0$. The details of the calculations are given in Secs.~S~I.B.1. and S~I.B.2. for 3D and 2D, respectively.
The final expressions can be summarized as follows:
\begin{eqnarray}
\label{app-def-lab-n-relat}
n &=& \frac{n_{\rm eq}}{\left(1-\beta^2\right)^{(d+1)/2}},\\
\label{app-def-lab-eps-relat}
\epsilon &=& \epsilon_{\rm eq} \frac{1+\beta^2/d}{\left(1-\beta^2\right)^{(d+3)/2}},\\
\label{app-def-lab-P-relat}
P &=& \epsilon \frac{1-\beta^2}{d+\beta^2} = \frac{\epsilon_{\rm eq}}{d}\frac{1}{\left(1-\beta^2\right)^{(d+1)/2}},\\
\label{app-def-lab-w-relat}
w &=& \epsilon+P= \epsilon_{\rm eq} \frac{d+1}{d}\frac{1}{\left(1-\beta^2\right)^{(d+3)/2}},\\
\label{app-def-lab-Pij-relat}
\Pi_{ij} &=& \frac{\epsilon}{d} \frac{1}{1+\beta^2/d} \left[\delta_{ij}\left(1-\beta^2\right) +(d+1)\frac{u_iu_j}{v_F^2}\right],
\end{eqnarray}
where $\beta=u/v_F$ and variables $n_{\rm eq}$ and $\epsilon_{\rm eq}$, are defined in the comoving reference frame presented in Sec.~S~I.A.

Naively, Eq.~(\ref{app-def-lab-n-relat}) suggests that the electron density is a function of the fluid velocity for a relativisticlike dispersion relation. Indeed, unlike truly relativistic fluids, there is no Lorentz contraction for the electrons in Dirac and Weyl materials ($v_F$ does not play the same role as $c$). Therefore, the volume of a fluid parcel remains the same regardless of its velocity and, consequently, the density of electrons should not change. To resolve the apparent paradox, one should assume that the dependence on $\mathbf{u}$ appears in the chemical potential $\mu$, which is present in $n_{\rm eq}$; see Eq.~(\ref{app-def-com-n}) for the 3D case.

\subsubsection{S I.B.1. 3D case}
\label{sec:app-def-lab-3D}

In this section, we consider the derivation of the key thermodynamic variables in the 3D case. The electric charge density reads as
\begin{eqnarray}
\label{app-def-lab-3D-n}
-en &=& -\sum_{\lambda}\sum_{\rm e,h} \frac{e}{(2\pi \hbar)^3} \int_0^{\infty} p^2 dp\, \int_{0}^{\pi} \sin{\theta} d\theta \int_{0}^{2\pi}d\varphi \frac{1}{1+e^{\left[v_F p\left(1- \beta\cos{\theta}\right) -\mu\right]/T}}
= - \frac{en_{\rm eq}}{2} \int_{-1}^{1} \frac{d\cos{\theta}}{\left(1-\beta \cos{\theta}\right)^3} \nonumber\\
&=& -\frac{en_{\rm eq}}{\left(1-\beta^2\right)^2},
\end{eqnarray}
where we rescaled momentum and used Eqs.~(\ref{app-def-n}) and (\ref{app-def-com-n}).

The energy density can be obtained along the same lines. It reads as
\begin{equation}
\label{app-def-lab-3D-eps}
\epsilon = \sum_{\lambda}\sum_{\rm e,h} \frac{1}{(2\pi \hbar)^3} \int_0^{\infty} v_Fp^3 dp\, \int_{0}^{\pi} \sin{\theta} d\theta \int_{0}^{2\pi}d\varphi \frac{1}{1+e^{\left[v_F p\left(1- \beta\cos{\theta}\right) -\mu\right]/T}} = \frac{\epsilon_{\rm eq}}{2} \int_{-1}^{1} \frac{d\cos{\theta}}{\left(1-\beta \cos{\theta}\right)^4}
=\epsilon_{\rm eq} \frac{1+\beta^2/3}{\left(1-\beta^2\right)^3},
\end{equation}
where we used Eqs.~(\ref{app-def-eps}) and (\ref{app-def-com-eps}).

Pressure is
\begin{eqnarray}
\label{app-def-lab-3D-P}
P &=& \sum_{\lambda}\sum_{\rm e,h} \frac{1}{(2\pi \hbar)^3} \int_0^{\infty} v_Fp^3 dp\, \int_{0}^{\pi} \sin{\theta} d\theta \int_{0}^{2\pi}d\varphi \sin^2{\theta} \cos^2{\varphi} \frac{1}{1+e^{\left[v_F p\left(1- \beta\cos{\theta}\right) -\mu\right]/T}}
= \frac{\epsilon_{\rm eq}}{4} \int_{-1}^{1} \frac{\sin^2{\theta} d\cos{\theta}}{\left(1-\beta \cos{\theta}\right)^4} \nonumber\\
&=& \frac{\epsilon_{\rm eq}}{3} \frac{1}{\left(1-\beta^2\right)^2}.
\end{eqnarray}

Finally, the enthalpy equals
\begin{equation}
\label{app-def-lab-3D-w}
w = \epsilon+P = \frac{4 \epsilon_{\rm eq}}{3}\frac{1}{\left(1-\beta^2\right)^{2}}.
\end{equation}

\subsubsection{S I.B.2. 2D case}
\label{sec:app-def-lab-2D}

Let us proceed now to the 2D case. The electric charge density reads as
\begin{equation}
\label{app-def-lab-2D-n}
-en = -\sum_{\lambda}\sum_{\rm e,h} \frac{e}{(2\pi \hbar)^2} \int_0^{\infty} p dp\, \int_{0}^{2\pi}d\varphi \frac{1}{1+e^{\left[v_F p\left(1- \beta\cos{\varphi}\right) -\mu\right]/T}}
= - \frac{en_{\rm eq}}{2\pi} \int_{0}^{2\pi} \frac{d\varphi}{\left(1-\beta \cos{\varphi}\right)^2} 
=- \frac{en_{\rm eq}}{\left(1-\beta^2\right)^{3/2}},
\end{equation}
where we used Eqs.~(\ref{app-def-n}) and (\ref{app-def-com-n-2D}).

The energy density is
\begin{equation}
\label{app-def-lab-2D-eps}
\epsilon = \sum_{\lambda}\sum_{\rm e,h} \frac{1}{(2\pi \hbar)^2} \int_0^{\infty} v_Fp^2 dp\, \int_{0}^{2\pi}d\varphi \frac{1}{1+e^{\left[v_F p\left(1- \beta\cos{\varphi}\right) -\mu\right]/T}}
= \frac{\epsilon_{\rm eq}}{2\pi} \int_{0}^{2\pi} \frac{d\varphi}{\left(1-\beta \cos{\varphi}\right)^3}
= \epsilon_{\rm eq} \frac{1+\beta^2/2}{\left(1-\beta^2\right)^{5/2}},
\end{equation}
where we used Eqs.~(\ref{app-def-eps}) and (\ref{app-def-com-eps-2D})

Pressure is given by
\begin{equation}
\label{app-def-lab-2D-P}
P = \sum_{\lambda}\sum_{\rm e,h} \frac{1}{(2\pi \hbar)^2} \int_0^{\infty} v_Fp^2 dp\, \int_{0}^{2\pi}d\varphi \sin^2{\varphi} \frac{1}{1+e^{\left[v_F p\left(1- \beta\cos{\varphi}\right) -\mu\right]/T}}
= \frac{\epsilon_{\rm eq}}{2\pi} \int_{0}^{2\pi} \frac{\sin^2{\varphi} d\varphi}{\left(1-\beta \cos{\varphi}\right)^3}
= \frac{\epsilon_{\rm eq}}{2} \frac{1}{\left(1-\beta^2\right)^{3/2}}.
\end{equation}

Combining Eqs.~(\ref{app-def-lab-2D-eps}) and (\ref{app-def-lab-2D-P}), we obtain the following expression for the enthalpy:
\begin{equation}
\label{app-def-lab-2D-w}
w = \epsilon+P = \frac{3\epsilon_{\rm eq}}{2}\frac{1}{\left(1-\beta^2\right)^{2}}.
\end{equation}

\section{S II. Solutions and characteristic equations}
\label{sec:app-sol}

The linearized hydrodynamic equations for a charged electron fluid are presented in the main text. For the sake of convenience, we present them here too. They read as
\begin{eqnarray}
\label{app-sol-eq-lin-1}
&&\left(\omega -2k_x u_0\right) u_1 +\frac{u_0}{w_0} \left(\omega -k_x u_0\right) w_1 -k_x \frac{v_F^2}{w_0} P_1 +i\frac{u_0}{\tau}\left(\frac{u_1}{u_0} +\frac{w_1}{w_0}-\frac{n_1}{n_0} \right) =-k_x\frac{e n_0 v_F^2}{w_0} \varphi_1,\\
\label{app-sol-eq-lin-2}
&&\left(\omega -k_x u_0\right) n_1 = k_x n_0 u_1,\\
\label{app-sol-eq-lin-3}
&&\omega \epsilon_1 -k_x u_0 w_1 -k_xw_0 u_1 -i\frac{1}{\tau}\frac{u_0^2}{v_F^2} w_0 \left(\frac{n_1}{n_0} +\frac{u_1}{u_0}\right) =-u_0 k_x en_0 \varphi_1,\\
\label{app-sol-eq-lin-4}
&&\mbox{3D:}\quad \varphi_1 = -\frac{4\pi e}{k^2} n_1 \quad \mbox{or} \quad \mbox{2D:}\quad \varphi_1 = -\frac{e}{C} n_1.
\end{eqnarray}
Here, we employed a conventional linear stability analysis where weak fluctuations are superimposed on top of a steady uniform flow determined by the fluid velocity $\mathbf{u}_0 \parallel \hat{\mathbf{x}}$, i.e.,
\begin{equation}
\label{app-sol-dn}
u_x(t,\mathbf{r}) =u_0+ u_1 e^{-i\omega t+i\mathbf{k}\cdot\mathbf{r}},
\end{equation}
and similar expressions for other variables $n$, $\varphi$, and $\epsilon$ were used. Further, $\omega$ and $\mathbf{k}=k_x \hat{\mathbf{x}}$ are the angular frequency and the wave vector of excitations, respectively. In the 2D system, we use the ``gradual channel" approximation~\cite{Shur:book,Dyakonov-Shur:1993} to determine the electric potential deviation $\varphi_1$, see Eq.~(\ref{app-sol-eq-lin-4}), where $C=\varepsilon/(4\pi L_g)$ is the capacitance per unit area, $\varepsilon$ is the dielectric constant of the substrate, and $L_g$ is the distance to the gate.

\subsection{S II.A. Boundary conditions}
\label{sec:app-sol-BC}

At the boundaries $x=0$ and $x=L$, we employ the standard Dyakonov-Shur boundary conditions~\cite{Dyakonov-Shur:1993} and fix temperature at one of the surfaces because of the added energy continuity equation. The boundary conditions read as
\begin{eqnarray}
\label{app-sol-BC-n}
n_1(x=0) &=& 0,\\
\label{app-sol-BC-J}
J_x(x=L) &\equiv& n_0u_1(x=L) +u_0n_1(x=L)= 0,\\
\label{app-sol-BC-T}
T_1(x=0) &=& 0.
\end{eqnarray}

We find it convenient to solve the hydrodynamic equations in terms of deviations $n_1$, $u_1$, and $\epsilon_1$. For this, we reexpress the boundary condition for temperature (\ref{app-sol-BC-T}) in terms of energy deviation $\epsilon_1$. By using Eqs.~(\ref{app-def-lab-n-relat}) and (\ref{app-def-lab-eps-relat}), we obtain
\begin{eqnarray}
\label{app-sol-BC-n1}
n_1 &=&  \frac{\left[\partial_{\mu} n_{\rm eq}\left(\mu(u),T(u)\right)\right]}{\left(1-\beta_0^2\right)^{(d+1)/2}} \mu_1(u) + \frac{\left[\partial_{T} n_{\rm eq}\left(\mu(u),T(u)\right)\right]}{\left(1-\beta_0^2\right)^{(d+1)/2}} T_1(u) +n_{\rm eq}\left(\mu(u_0),T(u_0)\right)\left[\partial_{u}\frac{1}{\left(1-\beta^2\right)^{(d+1)/2}}\right]_{u=u_0} u_1,\nonumber\\
\\
\label{app-sol-BC-eps1}
\epsilon_1 &=& \frac{1+\beta_0^2/d}{\left(1-\beta_0^2\right)^{(d+3)/2}} \left[\partial_{\mu} \epsilon_{\rm eq}\left(\mu(u),T(u)\right)\right] \mu_1(u) +\frac{1+\beta_0^2/d}{\left(1-\beta_0^2\right)^{(d+3)/2}} \left[\partial_{T} \epsilon_{\rm eq}\left(\mu(u),T(u)\right)\right] T_1(u) \nonumber\\
&+&\epsilon_{\rm eq}\left(\mu(u_0),T(u_0)\right)\left[\partial_{u}\frac{1+\beta^2/d}{\left(1-\beta^2\right)^{(d+3)/2}}\right]_{u=u_0}  u_1.
\end{eqnarray}
Here, we explicitly showed the dependence of $n_{\rm eq}$ and $\epsilon_{\rm eq}$ on temperature and the chemical potential. Recall that since the electron density does not depend on the fluid velocity, Eq.~(\ref{app-def-lab-n-relat}) means that the chemical potential and/or temperature should be functions of $u$.

By using Eqs.~(\ref{app-sol-BC-n}), (\ref{app-sol-BC-T}), and (\ref{app-sol-BC-n1}), we obtain
\begin{equation}
\label{app-sol-BC-mu1}
\mu_1(u)= -u_1 \left[\partial_{u} \frac{\left(1-\beta_0^2\right)^{(d+1)/2}}{\left(1-\beta^2\right)^{(d+1)/2}}\right]_{u=u_0} \frac{n_{\rm eq}\left(\mu(u_0),T(u_0)\right)}{\left[\partial_{\mu}n_{\rm eq}\left(\mu(u),T(u)\right)\right]}.
\end{equation}
The boundary condition for the energy density deviation reads as
\begin{eqnarray}
\label{app-sol-BC-eps1-BC}
\epsilon_1(x=0) &=& u_1 \Bigg\{\epsilon\left(\mu(u_0),T(u_0)\right)\left[\partial_{u}\frac{1+\beta^2/d}{\left(1-\beta^2\right)^{(d+3)/2}}\right]_{u=u_0}
- \left[\partial_{u}\frac{\left(1-\beta_0^2\right)^{(d+1)/2}}{\left(1-\beta^2\right)^{(d+1)/2}} \right]_{u=u_0} \frac{n\left(\mu(u_0),T(u_0)\right)}{\partial_{\mu}n_{\rm eq}\left(\mu(u),T(u)\right)} \nonumber\\ &\times&\frac{1+\beta_0^2/d}{\left(1-\beta_0^2\right)^{(d+3)/2}} \left[\partial_{\mu} \epsilon_{\rm eq}\left(\mu(u),T(u)\right)\right]\Bigg\} \nonumber\\
&=& \epsilon_0 u_1 \left\{\frac{\epsilon\left(\mu(u_0),T(u_0)\right)}{\epsilon_0} \left[\partial_{u}\frac{1+\beta^2/d}{\left(1-\beta^2\right)^{(d+3)/2}}\right]_{u=u_0}
-\frac{(d+1)^2}{d} \frac{1}{\left(1-\beta_0^2\right)^2} \frac{\beta_0}{v_F}\frac{\omega_p^2}{(v_sq_{\rm TF})^2}\right\} \nonumber\\
&=&\frac{u_1u_0}{v_s^2} \frac{d+1}{d^2} \epsilon_0 \left[1-(d+1)\left(1-\Lambda_p^2\right)\right].
\end{eqnarray}
Here,
\begin{equation}
\label{app-sol-3D-vs-def}
v_s = \frac{v_F}{\sqrt{d}}
\end{equation}
is the sound velocity in a relativisticlike gas with $d=2,3$ being the spatial dimension. In Eq.~(\ref{app-sol-BC-eps1-BC}), we use the following shorthand notation:
\begin{equation}
\label{app-sol-BC-Omega-def}
\Lambda_{p} =  \frac{\omega_p}{v_s q_{\rm TF}} = \frac{v_F n_0}{v_s \sqrt{w_0 (\partial_{\mu}n_0)}},
\end{equation}
where the Thomas-Fermi wave vector and the plasma frequency for a relativisticlike fluid are given by
\begin{eqnarray}
\label{app-sol-BC-qTF-def}
q_{\rm TF} &=& \sqrt{4\pi e^2 \left(\partial_{\mu} n_0\right)},\\
\label{app-sol-BC-omegap-def}
\omega_p &=& \sqrt{\frac{4\pi e^2 n_0^2 v_F^2}{w_0}}.
\end{eqnarray}

\begin{figure}[t]
\includegraphics[width=0.4\textwidth]{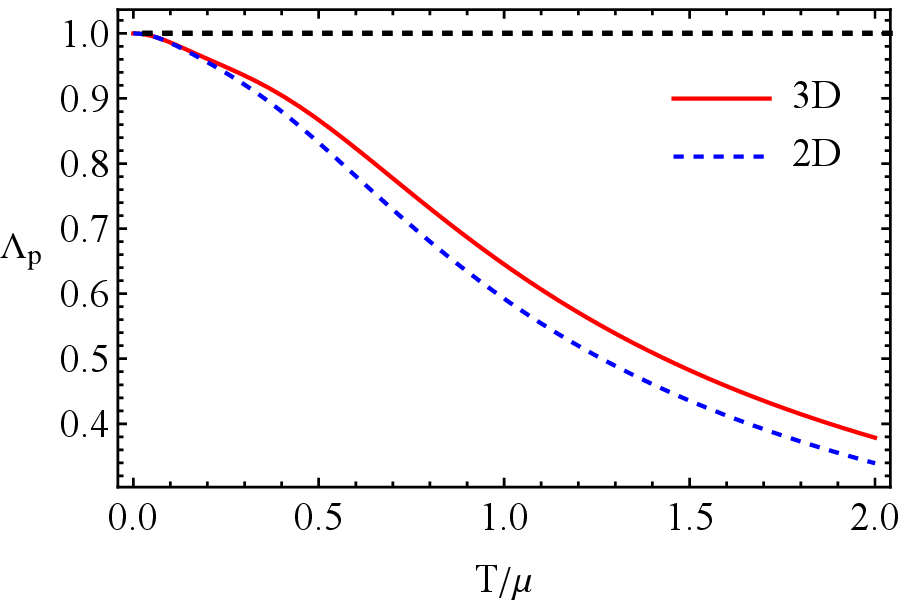}
\caption{Dependence of parameter $\Lambda_{p} = \omega_p/(v_s q_{\rm TF})$ on $T/\mu$ in 3D (red solid line) and 2D (blue dashed line).
}
\label{fig:SM-Lambda}
\end{figure}

The dependence of parameter $\Lambda_{p}$ on $T/\mu$ is shown in Fig.~\ref{fig:SM-Lambda}. As one can see, $\Lambda_{p}$ is nonnegative and always less than unity (i.e., $0<\Lambda_{p}<1$) for any nonvanishing temperature. It approaches $1$ from below only at zero temperature. Note, however, that the limiting case $T=0$ corresponds to the Fermi liquid regime rather than electron hydrodynamics.

\subsection{S II.B. 3D case}
\label{sec:app-sol-3D}

In this section, we discuss the solutions and the derivation of the characteristic equation for 3D systems. The solutions to the linearized equations of motion (\ref{app-sol-eq-lin-1}) through (\ref{app-sol-eq-lin-4}) at $\tau\to\infty$ are
\begin{eqnarray}
\label{app-sol-3D-sol-n}
\frac{n_1}{n_0} &=& \sum_{j}C_{j} e^{ik_{j} x},\\
\label{app-sol-3D-sol-u}
u_1 &=& \sum_{j}\left(\omega -u_0k_{j}\right) \frac{C_{j} e^{ik_{j} x}}{k_{j}},\\
\label{app-sol-3D-sol-phi}
\varphi_1 &=& -4\pi e n_0\sum_{j}\frac{C_{j} e^{ik_{j} x}}{k_j^2},\\
\label{app-sol-3D-sol-eps}
\frac{\epsilon_1}{\epsilon_0} &=& \frac{4}{9v_s^2\left[1+u_0^2/(9v_s^2)\right]}\sum_{j}\frac{C_je^{ik_jx}}{k_j \left\{\omega \left[1+u_0^2/(9v_s^2)\right] -4u_0k_j/3\right\}} \nonumber\\
&\times&\left[\omega_p^2 u_0\left(1+\frac{u_0^2}{9v_s^2}\right) +3v_s^2 \omega k_j \left(1-\frac{u_0^2}{9v_s^2}\right) -3v_s^2 u_0 k_j^2 \left(1-\frac{u_0^2}{9v_s^2}\right)\right].
\end{eqnarray}
Here, $\sum_{j}$ runs over all roots of the characteristic equation and $C_j$ are coefficients determined from the boundary conditions. As usual, system (\ref{app-sol-eq-lin-1}) through (\ref{app-sol-eq-lin-4}) has a nontrivial solution when its determinant vanishes. This leads to the following characteristic equation:
\begin{eqnarray}
\label{app-sol-3D-char-eq}
\left(\omega -u_0k_x\right)\left[\omega^2\left(1-\frac{u_0^2}{9v_s^2}\right) -\omega_p^2 \left(1-\frac{u_0^2}{3v_s^2}\right) -u_0 k_x\left(\frac{4}{3}\omega -u_0k_x\right)-v_s^2k_x^2\right]=0.
\end{eqnarray}
The roots of this equation read as:
\begin{eqnarray}
\label{app-sol-3D-k-pm}
k_{1,2} &=& \pm \frac{v_s}{1-u_0^2/v_s^2} \sqrt{\left(1-\frac{u_0^2}{3v_s^2}\right) \left[\omega^2-\omega_p^2 +\left(\omega_p^2-\frac{\omega^2}{3}\right) \frac{u_0^2}{v_s^2}\right]} -\frac{2}{3}\frac{\omega}{v_s} \frac{u_0/v_s}{1-u_0^2/v_s^2}
\approx \pm \frac{\sqrt{\omega^2-\omega_p^2}}{v_s} -\frac{2}{3} \frac{u_0 \omega}{v_s^2},\\
\label{app-sol-3D-k-e}
k_{3} &=& \frac{\omega}{u_0}.
\end{eqnarray}
We expanded up to the leading order in $u_0/v_s$ in the last expression in Eq.~(\ref{app-sol-3D-k-pm}).

The characteristic equation that determines the spectrum of collective excitations can be straightforwardly obtained by substituting solutions (\ref{app-sol-3D-sol-n}) through (\ref{app-sol-3D-sol-eps}) into the boundary conditions given in Eqs.~(\ref{app-sol-BC-n}), (\ref{app-sol-BC-J}), (\ref{app-sol-BC-eps1-BC}) and calculating the determinant of the corresponding system (with $C_1$, $C_2$, and $C_3$ being the independent variables now). The final result is, unfortunately, bulky. Therefore, we do not present it here.

To the linear order in $u_0$, the plasmon and entropy mode frequencies in 3D are given in the main text; see Eqs.~(23) and (25). The corresponding approximate results read as follows:
\begin{eqnarray}
\label{app-nums-3D-omega-plasmon-app}
\omega^{3D}_{\pm} &\approx&
\pm \sqrt{\omega_p^2 +\left[v_s\frac{\pi}{L} \left(l+\frac{1}{2}\right)\right]^2} + i\frac{2u_0}{3L} \left(3-2\Lambda_p^2\right)
-\frac{u_0^2}{3v_s^2}\frac{2\left(\omega_{\pm}^{(0)}\right)^4 -3\left(\omega_{\pm}^{(0)}\right)^2\omega_p^2 +\omega_p^4 -2v_s^2\omega_p^2/(3L^2)}{\omega_{\pm}^{(0)}\left[\left(\omega_{\pm}^{(0)}\right)^2-\omega_p^2\right]},\nonumber\\
\\
\label{app-nums-3D-omega-EW-app}
\omega^{3D}_{e} &\approx& \frac{2\pi l}{L} u_0 -i\frac{u_0 \omega_p}{v_s} -i \frac{u_0}{L} \ln{\left[\frac{3v_s^2}{8u_0^2 \left(1-\Lambda_p^2\right)}\right]},
\end{eqnarray}
where $l\in \mathds{Z}$ and $\omega_{\pm}^{(0)}=\pm \sqrt{\omega_p^2 +\left[v_s\pi \left(l+1/2\right)/L\right]^2}$. Note that we also included the terms quadratic in $u_0$ in Eq.~(\ref{app-nums-3D-omega-plasmon-app}). As is clear from the above expressions, the real part of the plasmon frequency has a weak dependence on the flow velocity $u_0$, which reduces the frequency.
This is also confirmed by the numerical results shown in Fig.~1 in the main text.

\subsection{S II.C. 2D case}
\label{sec:app-sol-2D}

In the 2D case, the solutions to the linearized equations of motion (\ref{app-sol-eq-lin-1}) through (\ref{app-sol-eq-lin-4}) at $\tau\to\infty$ are
\begin{eqnarray}
\label{app-sol-2D-sol-n}
\frac{n_1}{n_0} &=& \sum_{j}C_{j} e^{ik_{j} x},\\
\label{app-sol-2D-sol-u}
u_1 &=& \sum_{j}\left(\omega -u_0k_{j}\right) \frac{C_{j} e^{ik_{j} x}}{k_{j}} ,\\
\label{app-sol-2D-sol-phi}
\varphi_1 &=& -\frac{en_0}{C} \sum_{j}\frac{C_{j} e^{ik_{j} x}}{\omega -u_0k_{j}},\\
\label{app-sol-2D-sol-eps}
\frac{\epsilon_1}{\epsilon_0} &=& \frac{3}{4v_s^2 \left[1+u_0^2/(4v_s^2)\right]} \sum_{j} \frac{C_{j}e^{ik_{j} x}}{\omega \left[1-u_0^2/(4v_s^2)\right] -3u_0k_j/2} \left\{2v_s^2 \omega \left(1+\frac{u_0^2}{4v_s^2}\right) +u_0k_j \left[v_s^2\left(\xi -2\right) +\frac{u_0^2}{4v_s^2}\left(v_p^2 +v_s^2\right)\right]\right\}.\nonumber\\
\end{eqnarray}
Here, $\sum_{j}$ runs over the roots of the characteristic equation
\begin{eqnarray}
\label{DS-model-char-eq-2D}
\left(\omega -u_0k_x\right)\Bigg[\omega^2 \left(1 -\frac{u_0^2}{4v_s^2}\right) - u_0 k_x \omega
-v_s^2 k_x^2 \left(1-\frac{u_0^2}{v_s^2}\right) -v_s^2 k_x^2 \xi\left(1-\frac{u_0^2}{2v_s^2}\right) \Bigg]=0.
\end{eqnarray}
These roots are
\begin{eqnarray}
\label{app-sol-2D-k-pm}
k_{1,2} &=& -\frac{\omega u_0}{2} \frac{1}{v_p^2 -u_0^2/(2v_s^2) \left(v_p^2+v_s^2\right)} \pm \frac{\omega}{v_p^2 -u_0^2/(2v_s^2) \left(v_p^2+v_s^2\right)} \sqrt{1-\frac{u_0^2}{2v_s^2}}\sqrt{v_p^2 -\frac{u_0^2}{4v_s^2} \left(v_p^2+v_s^2\right)} \nonumber\\
&\approx&\pm\frac{\omega}{v_p} - \frac{u_0 \omega}{2v_p^2},\\
\label{app-sol-2D-k-e}
k_{3} &=& \frac{\omega}{u_0}.
\end{eqnarray}
Here, we introduced the following dimensionless parameter:
\begin{equation}
\label{app-sol-xi-def}
\xi = \frac{2e^2 n_0^2}{w_0 C},
\end{equation}
which determines the plasmon velocity $v_p=v_s \sqrt{1+\xi}$ in 2D.

The characteristic equation that determines the spectrum of collective excitations is obtained similarly to the 3D case and is also bulky.

To finalize this section, we present the results for plasmon and entropy mode frequencies in 2D. To the linear order in $u_0$, the approximate analytical expressions for the plasmons and entropy waves are given in Eqs.~(24) and (26) in the main text. Including the terms quadratic in $u_0$ in the plasmon dispersion relation, we have
\begin{eqnarray}
\label{app-nums-2D-omega-plasmon-app}
\omega^{2D}_{\pm} &\approx& \pm \frac{\pi}{L}\left(l+\frac{1}{2}\right) v_p +i\frac{u_0}{2L} \left(4-3\Lambda_p^2\right) \mp\frac{\pi}{L}\left(l+\frac{1}{2}\right)\frac{u_0^2}{8v_p} \frac{3v_s^2+v_p^2}{v_s^2},\\
\label{app-nums-2D-omega-EW-app}
\omega^{2D}_{e} &\approx& \frac{2\pi l}{L} u_0 -i \frac{u_0}{L} \ln{\left[\frac{2v_p^2}{3u_0^2 \left(1-\Lambda_p^2\right)}\right]},
\end{eqnarray}
where $l\in \mathds{Z}$. As in 3D, the frequency of plasmons is reduced at $u_0\neq0$.

\begin{figure*}[t]
\centering
\subfigure[]{\includegraphics[height=0.3\textwidth]{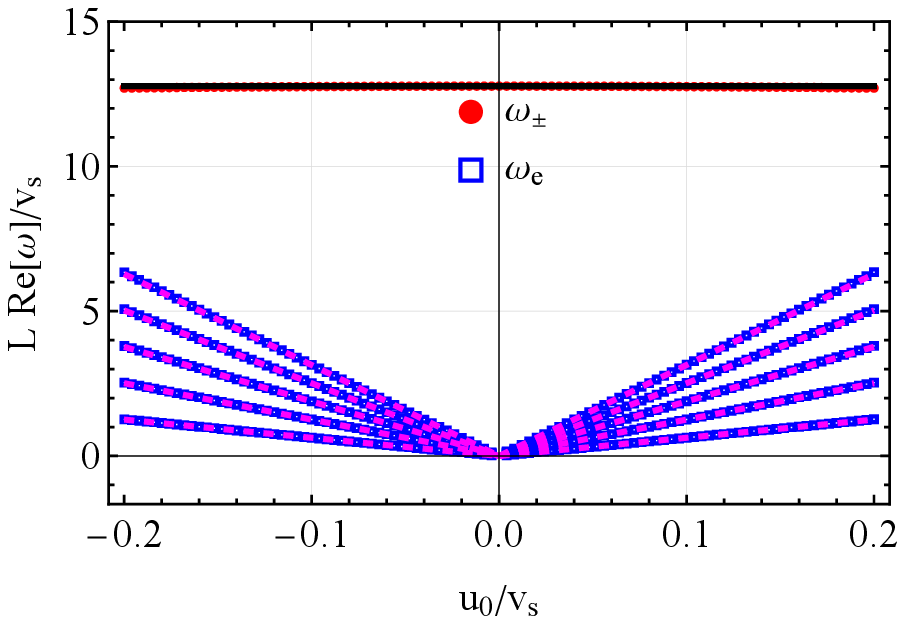}}
\hspace{0.05\textwidth}
\subfigure[]{\includegraphics[height=0.3\textwidth]{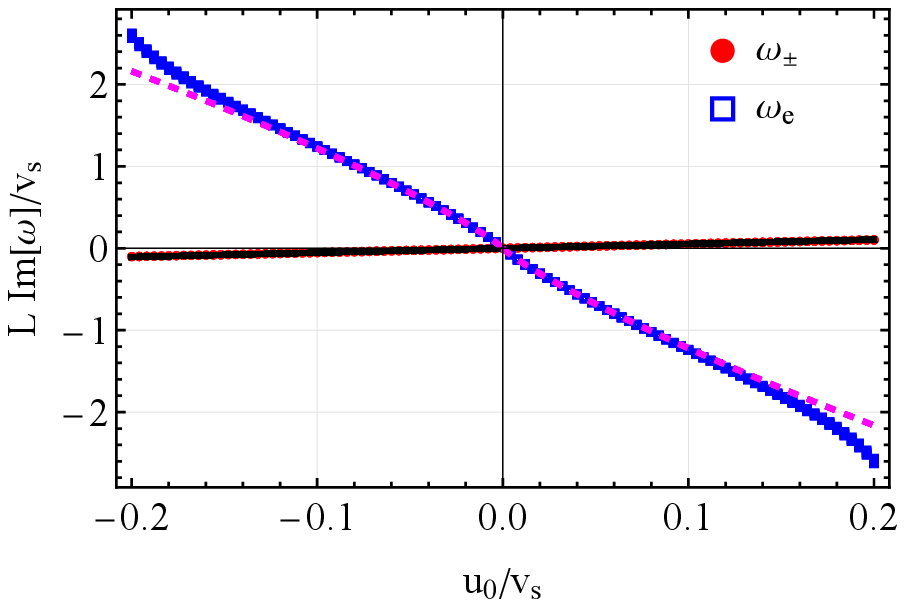}}
\vspace{-0.5cm}
\caption{
The real (panel (a)) and imaginary (panel (b)) parts of frequency as a function of velocity $u_0$ in the 2D case. We show only the lowest five branches of numerical and approximate results. Black solid lines correspond to the approximate expression (\ref{app-nums-2D-omega-plasmon-app}). The approximate relation (\ref{app-nums-2D-omega-EW-app}) is shown by magenta dashed lines. We fixed $v_s\approx7.8\times10^{7}~\mbox{cm/s}$, $v_{p}\approx 8.1 \, v_s$, and took the limit $\tau\to\infty$.
}
\label{fig:DS-inst-2D}
\end{figure*}

Our numerical and approximate results for the plasmon and entropy mode frequencies are given in Fig.~\ref{fig:DS-inst-2D}. As in the 3D case shown in the main text, we present only the lowest five branches of the numerical and approximate analytical results. We use the numerical parameters listed in the main text. There is a good agreement between the numerical results and the approximate expressions (\ref{app-nums-2D-omega-plasmon-app}) and (\ref{app-nums-2D-omega-EW-app}). As expected, plasmons have nonzero frequencies at $u_0\to0$ and the solutions for the entropy modes vanish at this limit. In addition, the growth rate of the instability is much larger for the entropy waves than for plasmons. Therefore, as in 3D, the entropy wave instability should be more pronounced than the Dyakonov-Shur one for the same flow velocities.

\section{S III. Numerical results for oscillating variables}
\label{sec:app-nums}

In this section, we present our results for oscillating variables in 2D and 3D cases.

\subsection{S III.A. 3D case}
\label{sec:app-nums-3D}

We plot the normalized hydrodynamic variables in the 3D case
\begin{equation}
\label{app-nums-3D-X1-def}
X_1=\left\{\frac{n_1}{n_0}, \frac{\epsilon_1}{\epsilon_0}, \frac{u_1}{v_s}, \frac{j_1}{n_0 v_s}, \frac{j_1^{\epsilon}}{\epsilon_0 v_s}\right\}
\end{equation}
for the entropy waves and plasmons in Figs.~\ref{fig:Instability-ref-3D-entropy} and \ref{fig:Instability-ref-3D-plasmon}, respectively. We use $L=10\, v_s/\omega_p$, $\Lambda_p \approx 0.98$, and $u_0\approx 0.1\,v_s$. All variables are normalized to the maximal values of $|n_1|/n_0$. The entropy modes are localized near the surface $x=L$ but show noticeable oscillations for larger $\mbox{Re}\left[\omega\right]$, see Figs.~\ref{fig:Instability-ref-3D-entropy}(c) and \ref{fig:Instability-ref-3D-entropy}(d). It is clear that the entropy modes are characterized by noticeable oscillations of the charge and energy densities. In agreement with our discussion in the main text, relative oscillations of the fluid velocity are small.

Unlike the entropy modes, the plasmon modes shown in several panels of Fig.~\ref{fig:Instability-ref-3D-plasmon} are delocalized. Furthermore, the charge and energy densities are no longer dominant there. Indeed, oscillations of the fluid velocity have almost the same magnitude as those for the densities. In addition, by comparing Figs.~\ref{fig:Instability-ref-3D-entropy} and \ref{fig:Instability-ref-3D-plasmon}, we notice that while the charge and energy densities oscillate in antiphase for the entropy mode, oscillations are in-phase for plasmons.

\subsection{S III.B. 2D case}
\label{sec:app-nums-2D}

\begin{figure}[t]
\centering
\subfigure[]{\includegraphics[height=0.25\textwidth]{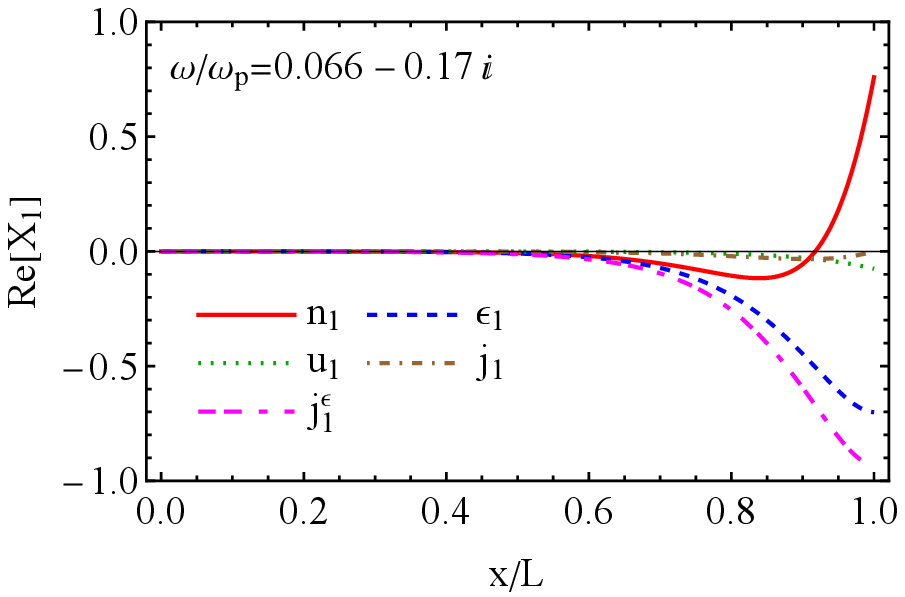}}
\hspace{0.05\textwidth}
\subfigure[]{\includegraphics[height=0.25\textwidth]{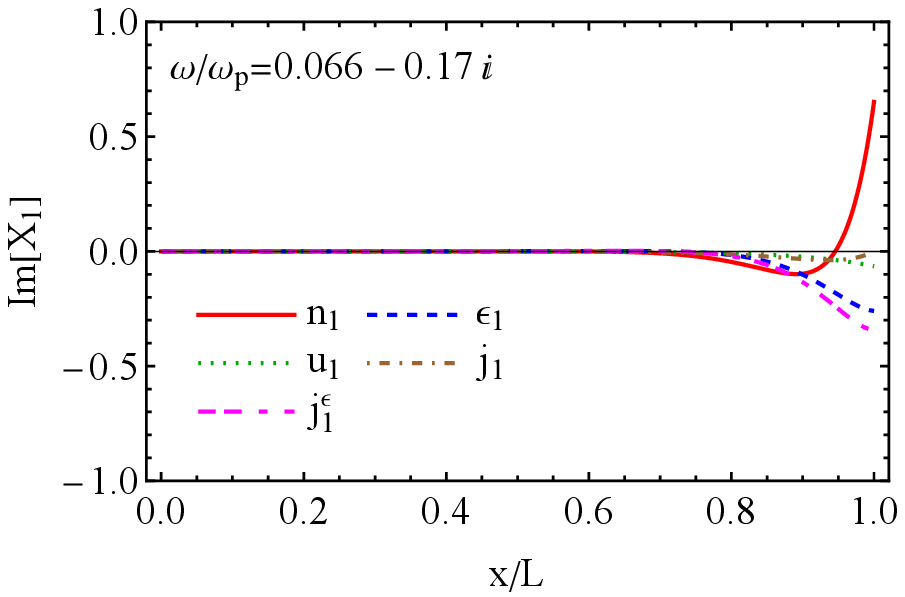}}
\hspace{0.05\textwidth}
\subfigure[]{\includegraphics[height=0.25\textwidth]{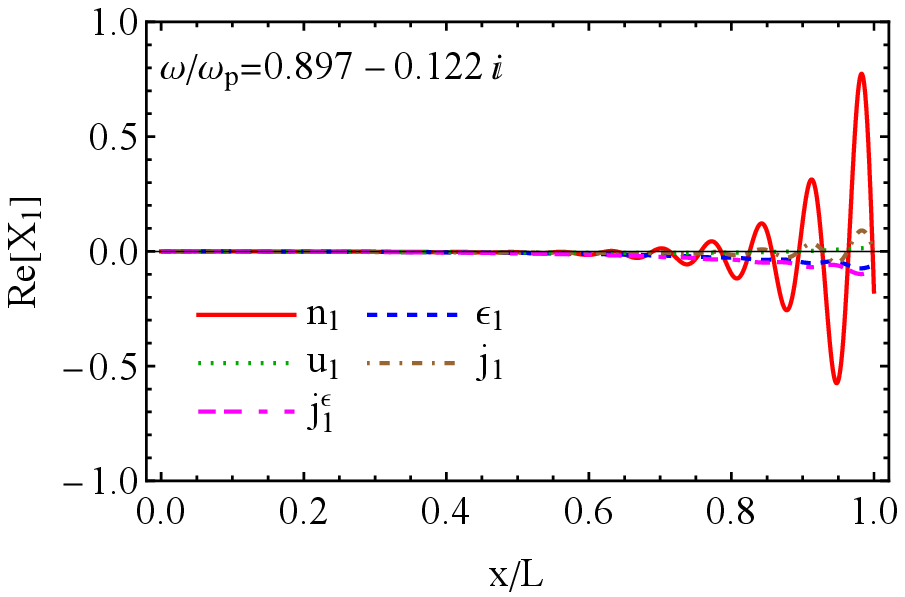}}
\hspace{0.05\textwidth}
\subfigure[]{\includegraphics[height=0.25\textwidth]{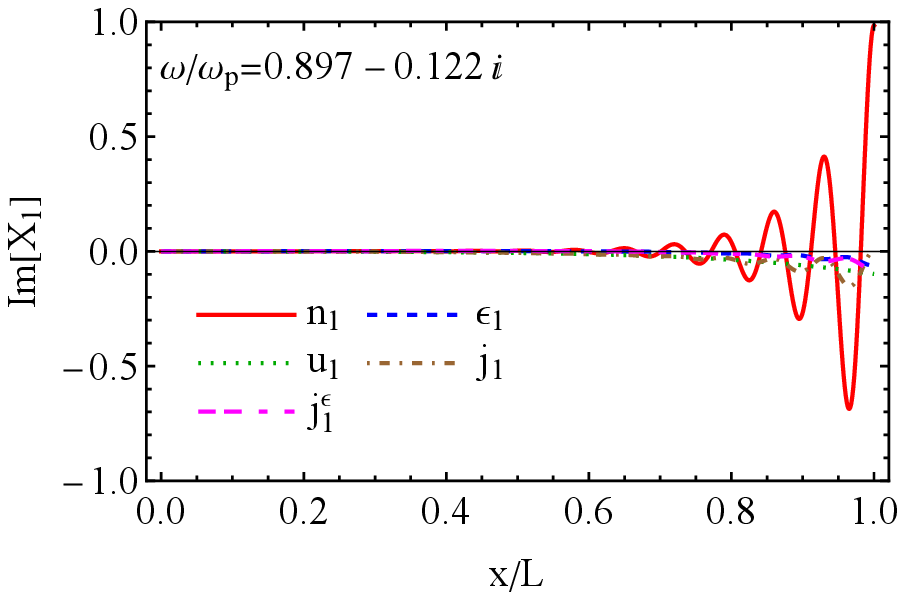}}
\vspace{-0.5cm}
\caption{The real and imaginary parts of the hydrodynamic variables $X_1$ defined in Eq.~(\ref{app-nums-3D-X1-def}) for a few branches of the entropy waves in 3D. In all panels, we set $L=10\, v_s/\omega_p$, $\Lambda_p \approx 0.98$, and $u_0\approx 0.1\,v_s$.}
\label{fig:Instability-ref-3D-entropy}
\end{figure}

\begin{figure}[t]
\centering
\subfigure[]{\includegraphics[height=0.25\textwidth]{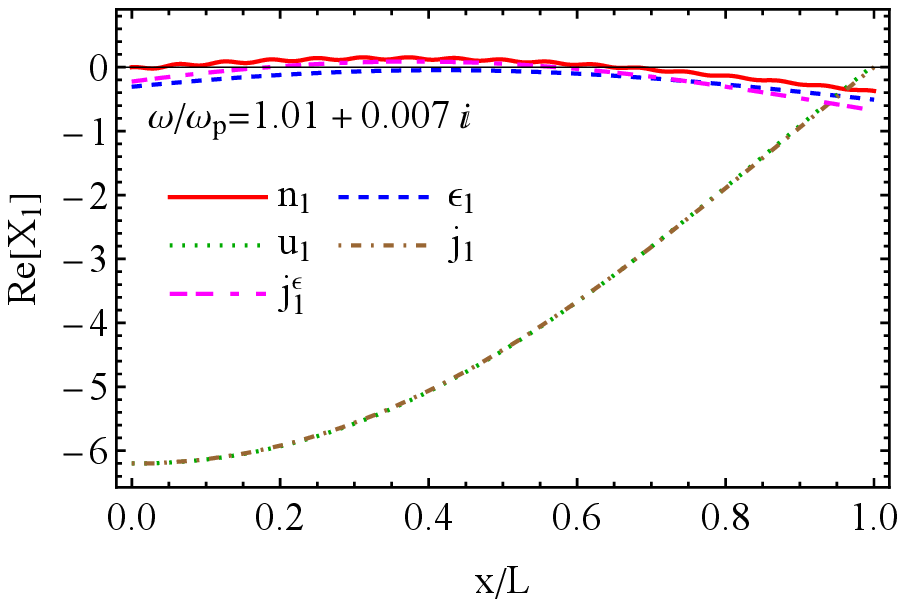}}
\hspace{0.05\textwidth}
\subfigure[]{\includegraphics[height=0.25\textwidth]{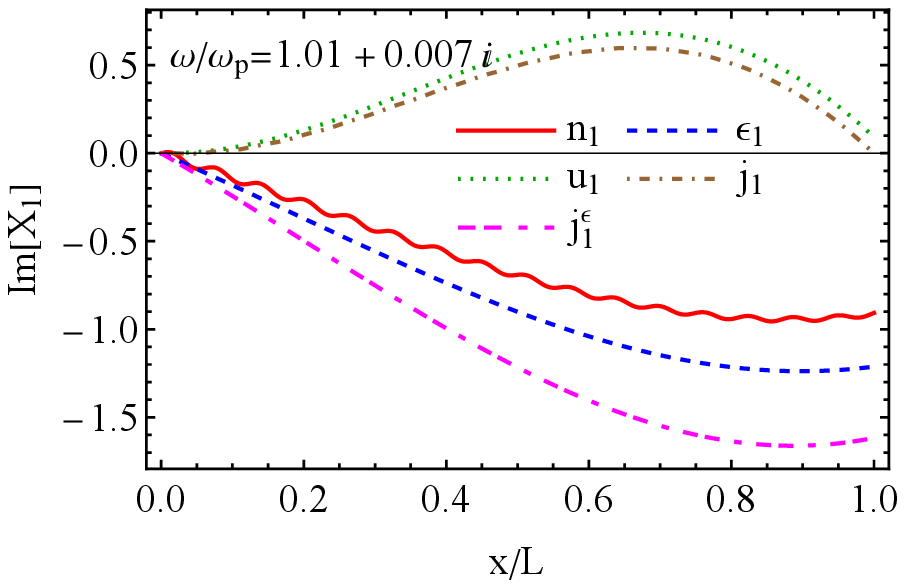}}
\hspace{0.05\textwidth}
\subfigure[]{\includegraphics[height=0.25\textwidth]{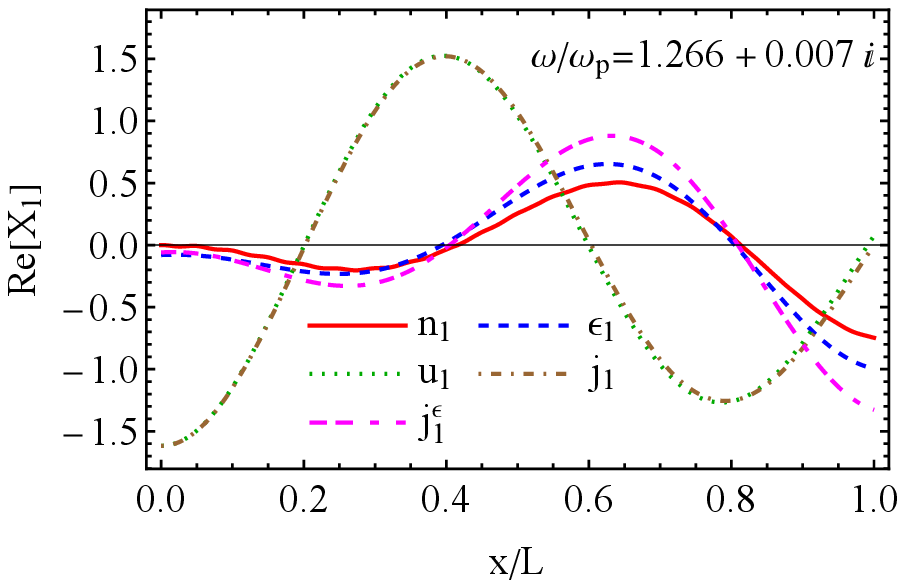}}
\hspace{0.05\textwidth}
\subfigure[]{\includegraphics[height=0.25\textwidth]{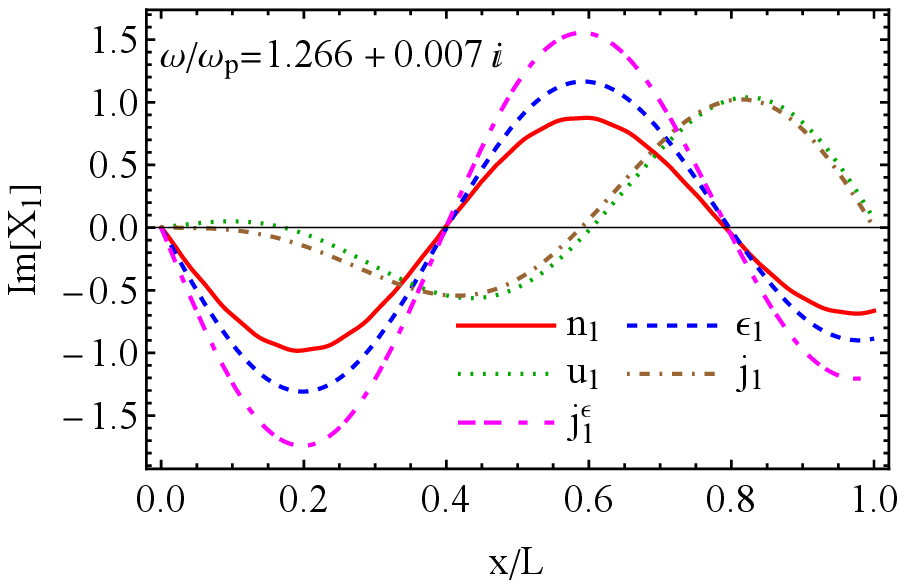}}
\vspace{-0.5cm}
\caption{The real and imaginary parts of the hydrodynamic variables $X_1$ defined in Eq.~(\ref{app-nums-3D-X1-def}) for a few  plasmon branches in 3D. In all panels, we set $L=10\, v_s/\omega_p$, $\Lambda_p \approx 0.98$, and $u_0\approx 0.1\,v_s$.}
\label{fig:Instability-ref-3D-plasmon}
\end{figure}

\begin{figure}[t]
\centering
\subfigure[]{\includegraphics[height=0.25\textwidth]{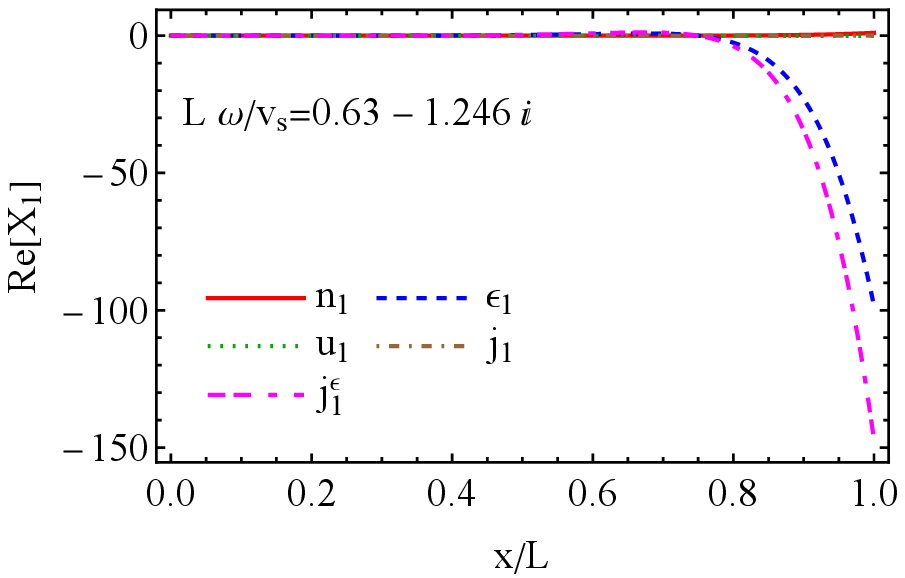}}
\hspace{0.05\textwidth}
\subfigure[]{\includegraphics[height=0.25\textwidth]{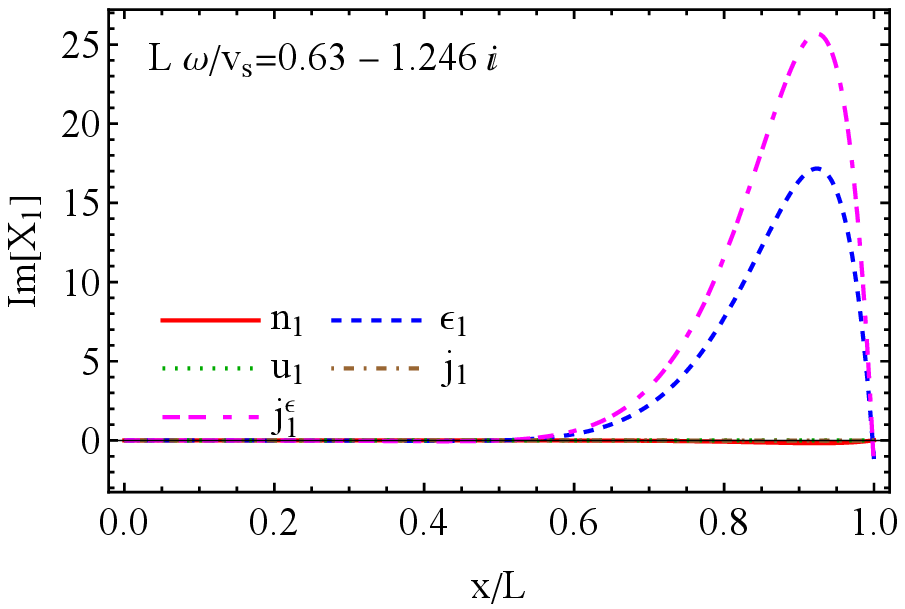}}
\hspace{0.05\textwidth}
\subfigure[]{\includegraphics[height=0.25\textwidth]{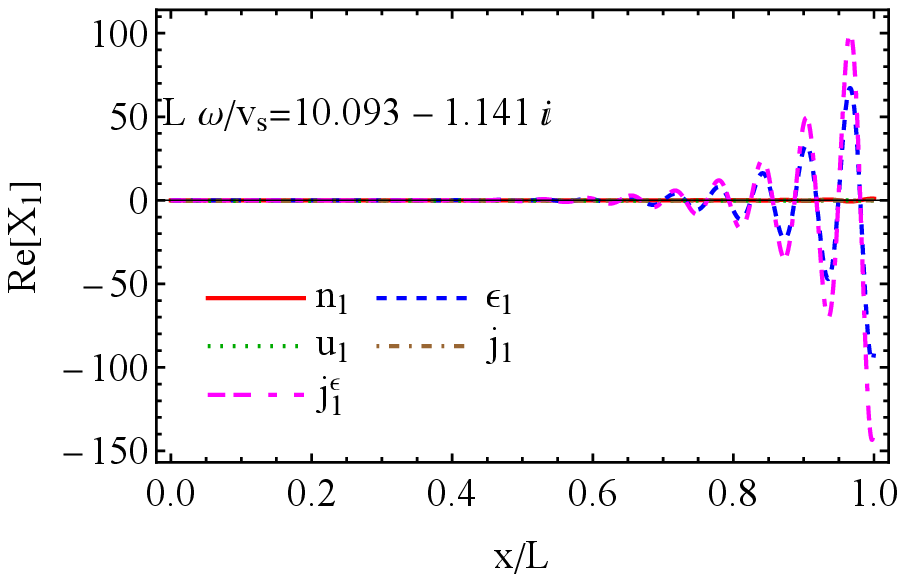}}
\hspace{0.05\textwidth}
\subfigure[]{\includegraphics[height=0.25\textwidth]{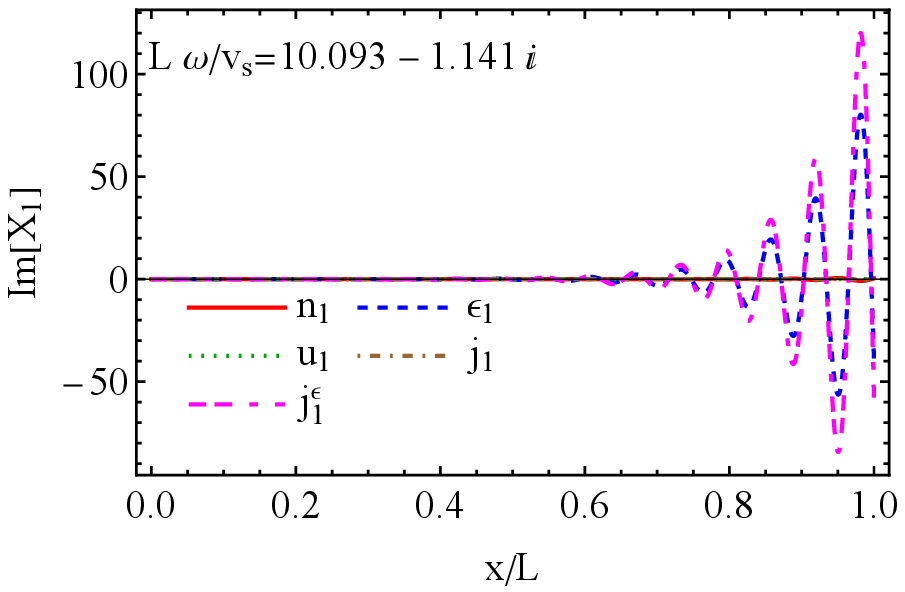}}
\vspace{-0.5cm}
\caption{The real and imaginary parts of the hydrodynamic variables $X_1$ defined in Eq.~(\ref{app-nums-3D-X1-def}) for a few branches of the entropy waves in 2D. In all panels, we set $\xi\approx 65.1$, $v_{p}\approx 8.1 \, v_s$, $\Lambda_p \approx 0.99$, $\tau\to\infty$, and $u_0\approx 0.1\,v_s$.}
\label{fig:Instability-ref-2D-entropy}
\end{figure}

\begin{figure}[t]
\centering
\subfigure[]{\includegraphics[height=0.25\textwidth]{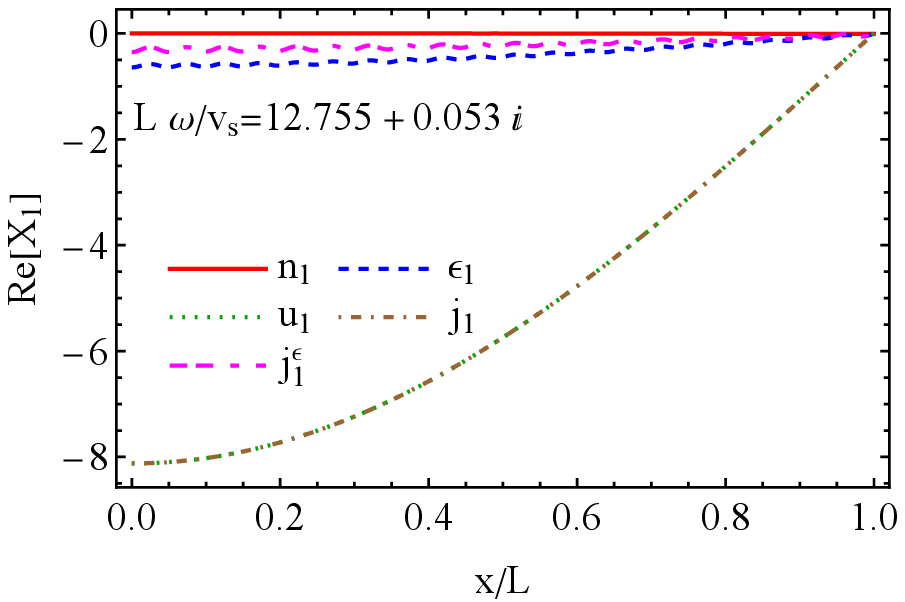}}
\hspace{0.05\textwidth}
\subfigure[]{\includegraphics[height=0.25\textwidth]{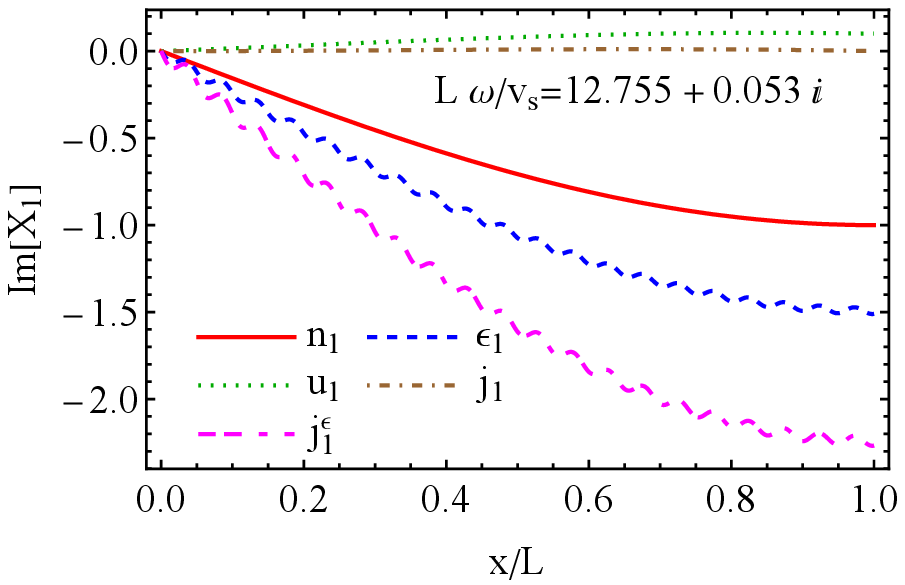}}
\hspace{0.05\textwidth}
\subfigure[]{\includegraphics[height=0.25\textwidth]{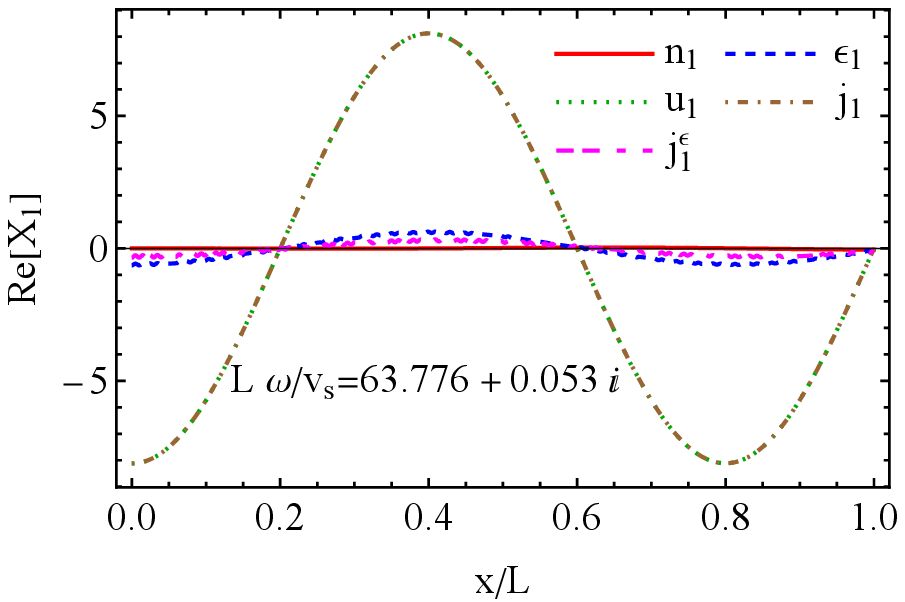}}
\hspace{0.05\textwidth}
\subfigure[]{\includegraphics[height=0.25\textwidth]{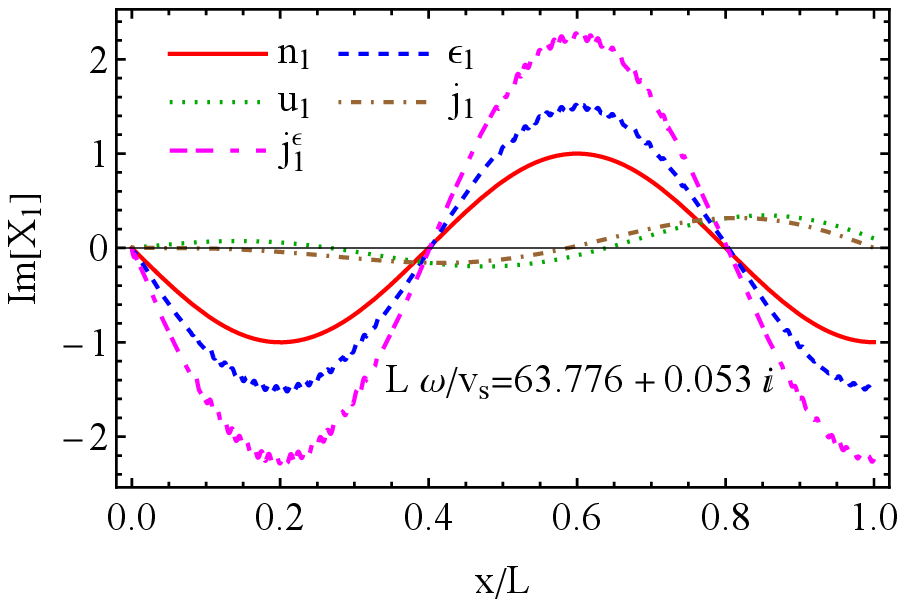}}
\vspace{-0.5cm}
\caption{The real and imaginary parts of the hydrodynamic variables $X_1$ defined in Eq.~(\ref{app-nums-3D-X1-def}) for a few plasmon branches in 2D. In all panels, we set $\xi\approx 65.1$, $v_{p}\approx 8.1 \, v_s$, $\Lambda_p \approx 0.99$, $\tau\to\infty$, and $u_0\approx 0.1\,v_s$.}
\label{fig:Instability-ref-2D-plasmon}
\end{figure}

Let us now proceed to the 2D relativisticlike case, e.g., graphene. We present the normalized hydrodynamic variables $X_1$ defined in Eq.~(\ref{app-nums-3D-X1-def}) for the entropy waves and plasmons in Figs.~\ref{fig:Instability-ref-2D-entropy} and \ref{fig:Instability-ref-2D-plasmon}, respectively, at $\xi\approx 65.1$, $v_{p}\approx 8.1 \, v_s$, $\Lambda_p \approx 0.99$, and $u_0\approx 0.1\,v_s$. All variables are normalized to the maximal $|n_1|/n_0$. As in 3D, the entropy modes are localized near the surface $x=L$ and show noticeable oscillations for larger $\mbox{Re}\left[\omega\right]$. On the other hand, the energy density oscillations are significantly larger than those for the charge density, cf. Figs.~\ref{fig:Instability-ref-3D-entropy} and \ref{fig:Instability-ref-2D-entropy}. Hydrodynamic variables for the plasmon modes are qualitatively similar to those in 3D, cf. Figs.~\ref{fig:Instability-ref-3D-plasmon} and \ref{fig:Instability-ref-2D-plasmon}.

\section{S IV. Dissipation effects}
\label{sec:app-dissipation}

Let us estimate the role of dissipative effects in the Dyakonov-Shur and entropy wave instabilities. As we discussed in the main text, the suppression of the plasmon instability can be roughly described by replacing~\cite{Dyakonov-Shur:1993}
$\omega\to\omega-i/\tau -i \eta_{\rm kin} \pi^2/L^2$, where $\eta_{\rm kin}$ is the kinematic viscosity. This leads to the following minimal flow velocities needed to overcome the dissipative effects:
\begin{eqnarray}
\label{app-dissipation-criterion-3D}
\mbox{3D:}\quad |u_0| &\gtrsim& \frac{3}{2} \frac{L}{3-2\Lambda_{p}^2}\left[\frac{1}{\tau} +\eta_{\rm kin} \left(\frac{\pi}{L}\right)^2\right],\\
\label{app-dissipation-criterion-2D}
\mbox{2D:}\quad |u_0| &\gtrsim& \frac{2L}{4-3\Lambda_{p}^2}\left[\frac{1}{\tau} +\eta_{\rm kin} \left(\frac{\pi}{L}\right)^2\right].
\end{eqnarray}
It is notable that while the small size of the system is beneficial for overcoming the momentum dissipation $\propto 1/\tau$, the contribution of viscosity rises for small $L$. Therefore, the Dyakonov-Shur instability is realized in a limited parameter range.

According to our analysis in Sec.~III,
oscillations of the fluid velocity are suppressed in the entropy wave instability. Therefore, the effects of momentum dissipation and viscosity are less pronounced. Phenomenologically, the dissipation effects (e.g., the intrinsic conductivity and diffusion) for the entropy wave instability can be described by introducing an effective relaxation time $\tau_{\rm eff}$. Then, by using Eqs.~(\ref{app-nums-3D-omega-EW-app}) and (\ref{app-nums-2D-omega-EW-app}), we estimate the following thresholds:
\begin{eqnarray}
\label{app-dissipation-criterion-EWI-3D}
\mbox{3D:}\quad |u_0| &\gtrsim& \frac{v_s}{\omega_p \tau_{\rm eff}},\\
\label{app-dissipation-criterion-EWI-2D}
\mbox{2D:}\quad |u_0| &\gtrsim& \frac{L}{\tau_{\rm eff}}.
\end{eqnarray}
Here, we assumed that $\omega_p L/v_s\gg1$ in 3D and neglected the logarithmic corrections. The dissipative effects for the entropy wave are negligible when $\omega_p \tau_{\rm eff}\gg1$ in 3D and $v_s\tau_{\rm eff}\gg L$ in 2D. By comparing Eqs.~(\ref{app-dissipation-criterion-3D}) and (\ref{app-dissipation-criterion-EWI-3D}) in 3D, we see that the entropy wave instability can be easier to achieve than the Dyakonov-Shur one.

\color{black}

\bibliography{library-short}